\documentclass{aa}  

\usepackage{graphicx}
\usepackage{txfonts}
\usepackage{xcolor}
\usepackage{xspace}
\usepackage{subfig}
\usepackage{balance}
\usepackage{ulem}

\usepackage{hyperref}
\hypersetup{colorlinks=true,citecolor=blue,linkcolor=blue,urlcolor=blue}

\newcommand{\tempotwo}{\texttt{TEMPO2}}
\newcommand{\psrchive}{\texttt{PSRCHIVE}}
\newcommand{\dspsr}{\texttt{DSPSR}}
\newcommand{\pinta}{\texttt{pinta}}
\newcommand{\rficlean}{\texttt{RFIClean}}
\newcommand{\dmcalc}{\texttt{DMcalc}}

\newcommand{\methodone}{\texttt{METHOD1}}
\newcommand{\methodtwo}{\texttt{METHOD2}}
\newcommand{\bandthree}{\textsl{BAND3}}
\newcommand{\bandfour}{\textsl{BAND4}}
\newcommand{\bandfive}{\textsl{BAND5}}

\begin{document}

\title{High precision measurements of interstellar dispersion measure with the upgraded GMRT}

   \author{M. A. Krishnakumar
          \inst{1}\thanks{E-mail: kkma@physik.uni-bielefeld.de},
          P. K. Manoharan\inst{2},
          Bhal Chandra Joshi\inst{3},
          Raghav Girgaonkar\inst{4},
          Shantanu Desai\inst{4},
          Manjari Bagchi\inst{5,6},
          K. Nobleson\inst{7},
          Lankeswar Dey\inst{8},
          Abhimanyu Susobhanan\inst{8},
          Sai Chaitanya Susarla\inst{9},
          Mayuresh P. Surnis\inst{10},
          Yogesh Maan\inst{11},
          A. Gopakumar\inst{8},
          Avishek Basu\inst{10},
          Neelam Dhanda Batra\inst{7,12},
          Arpita Choudhary\inst{5},
          Kishalay De\inst{13},
          Yashwant Gupta\inst{3},
          Arun Kumar Naidu\inst{14},
          Dhruv Pathak\inst{5,6},
          Jaikhomba Singha\inst{15},
          T. Prabu\inst{16}
          }

   \institute{
        Fakult{\"a}t f{\"u}r Physik, Universit{\"a}t Bielefeld, Postfach 100131, 33501 Bielefeld, Germany
        \and
        Arecibo Observatory, University of Central Florida, Arecibo, PR 00612, USA
        \and
        National Centre for Radio Astrophysics, Tata Institute of Fundamental Research, Ganeshkhind, Pune 411007, Maharashtra, India
        \and
        Department of Physics, Indian Institute of Technology Hyderabad, Kandi, Telangana 502285, India
        \and
        The Institute of Mathematical Sciences,  C. I. T. Campus, Tharamani, Chennai 600113, Tamil Nadu, India
        \and
        Homi Bhabha National Institute, Training School Complex, Anushakti Nagar, Mumbai 400094, Maharashtra, India
        \and
        Department of Physics, BITS Pilani Hyderabad Campus, Hyderabad 500078, Telangana, India 
        \and
        Department of Astronomy and Astrophysics, Tata Institute of Fundamental Research, Dr. Homi Bhabha Road, Mumbai 400005, Maharashtra, India
        \and
        Indian Institute of Science Education and Research Thiruvananthapuram, Vithura,  Kerala 695551, India
        \and
        Jodrell Bank Centre for Astrophysics, University of Manchester, Oxford Road, Manchester, M13 9PL, UK
        \and
        ASTRON, the Netherlands Institute for Radio Astronomy, Postbus 2, 7990 AA, Dwingeloo, The Netherlands
        \and
        Department of Physics, Indian Institute of Technology Delhi, New Delhi-110016, India
        \and
        Cahill Center for Astrophysics, California Institute of Technology, 1200 E. California Blvd. Pasadena, CA 91125, USA
        \and
        University of Oxford, Sub-Department of Astrophysics, Denys Wilkinson Building, Keble Road, Oxford, OX1 3RH, United Kingdom
        \and
        Department of Physics, Indian Institute of Technology Roorkee, Roorkee 247667, Uttarakhand, India
        \and
        Raman Research Institute, Bengaluru 560080, Karnataka, India
        }

   \date{Received XXX XX, XXXX; accepted YYY YY, YYYY}

\titlerunning{InPTA: Precision DM estimates with uGMRT}
\authorrunning{Krishnakumar et al. } 
  \abstract
   {Pulsar radio emission undergoes dispersion due to the presence of free electrons in the 
   interstellar medium (ISM). The dispersive delay in the arrival time of the pulsar signal 
   changes over time due to the varying ISM electron column density along the line of sight. 
   Accurately correcting for this delay is crucial for the detection of nanohertz gravitational 
   waves using pulsar timing arrays.
  }
   {We aim to demonstrate the precision in the measurement of the dispersion delay achieved 
   by combining 400$-$500\,MHz (\bandthree{}) wide-band data with those at 1360$-$1460 MHz 
   (\bandfive{}) observed using the upgraded GMRT, employing two different template alignment 
   methods. 
   }
   {To estimate the high precision dispersion measure (DM), we measure high precision 
   times-of-arrival (ToAs) of pulses using carefully generated templates and the currently 
   available pulsar timing techniques. We use two different methods for aligning the 
   templates across frequency to obtain ToAs over multiple sub-bands and therefrom measure 
   the DMs. We study the effects of these two different methods on the measured DM values 
   in detail.
   }
   {We present in-band and inter-band DM estimates of four pulsars over the timescale of 
   a year using two different template alignment methods. The DMs obtained using both these 
   methods show only subtle differences for PSRs J1713+0747 and J1909$-$3744. A considerable 
   offset is seen in the DM of PSRs J1939+2134 and J2145$-$0750 between the two methods. 
   This could be due to the presence of scattering in the former and profile evolution in 
   the latter. We find that both methods are useful but could have a systematic offset 
   between the DMs obtained. Irrespective of the template alignment methods followed, the 
   precision on the DMs obtained is about $10^{-3}$ pc~cm$^{-3}$ using only \bandthree{} and 
   $10^{-4}$ pc~cm$^{-3}$ after combining data from \bandthree{} and \bandfive{} of the uGMRT. 
   In a particular result, we detected a DM excess of about $5\times10^{-3}$ pc~cm$^{-3}$ on 
   24 February 2019 for PSR J2145$-$0750. This excess appears to be due to the interaction 
   region created by fast solar wind from a coronal hole and a coronal mass ejection (CME) 
   observed from the Sun on that epoch. A detailed analysis of this interesting event is 
   presented.
   }
   {}


   \maketitle
%

\section{Introduction}



Pulsars are rotating neutron stars that emit broadband radiation received as pulsed signals 
by the observers. The pulsar radiation reaches the observer after propagating through the 
ionised interstellar medium (IISM), which disperses the pulsed signal, thereby delaying the 
times of arrival (ToAs) of pulses as a function of the observing frequency 
\citep{LorimerKramer2004_handbook}. This dispersion delay is directly proportional to the 
integrated column density of free electrons in the IISM, usually referred to as the dispersion 
measure (DM), and inversely proportional to the square of the observing frequency ($\nu$). 
Precise measurements of the DM can therefore be made by measuring the pulse ToAs simultaneously 
at different observing frequencies \citep[e.g.][]{Backer1996,Ahuja2005,Ahuja2007}.

The DM of a pulsar can vary with time due to a number of factors, which include the relative 
motion of the pulsar with respect to the observer, solar wind, the terrestrial ionosphere, 
and the dynamical nature of the IISM. Typical DM variations observed in pulsars range from 
10$^{-3}$ -- 10$^{-5}$ pc~cm$^{-3}$ \citep{Ujjwal2012,ng12.5yr,DVT+20}. If these variations 
are not accounted for, systematic errors of the order of 1~$\mu$s or more can arise while 
correcting for the DM delay to generate infinite-frequency ToAs in the Solar System barycentre 
(SSB) frame \citep{Hobbs2006,Edwards2006}. Such unaccounted systematics have the potential 
to degrade the ability of millisecond pulsars (MSPs) to act as very accurate celestial clocks 
\citep{Hobbs2020_IPTA_clock}. The technique of pulsar timing that creates such celestial 
clocks requires  us to model and correctly characterise the pulse propagation effects 
\citep{Edwards2006}. This technique is crucial for the rapidly maturing pulsar timing array 
(PTA) efforts to detect nanohertz gravitational waves 
\citep[GWs;][]{FosterBacker1990,Arzoumanian2020}. Pulsar timing arrays pursue the timing of 
tens of MSPs to detect mainly a stochastic nanohertz GW background due to an ensemble of 
merging supermassive black hole binaries \citep{Burke-Spolaor2019}.

There are three established PTA efforts: are the Parkes Pulsar Timing Array
\citep[PPTA;][]{Hobbs2013,Kerr2020}, the European Pulsar Timing Array 
\citep[EPTA;][]{KramerChampion2013,Desvignes2016}, and the North American Nanohertz Observatory 
for Gravitational Waves \citep[NANOGrav;][]{McLaughlin2013,ng12.5yr}. In addition, PTA efforts 
are gathering pace in India under the auspices of the Indian Pulsar Timing Array 
\citep[InPTA;][]{Joshi2018}. The International Pulsar Timing Array (IPTA) consortium combines 
data and resources from various PTA efforts to enable a faster detection of nanohertz GWs 
\citep{Perera2019}. It should be noted that high precision DM measurements are essential for 
reaching the desired sensitivities of existing PTAs as precise pulse ToA estimates depend on 
accurate DM measurements. While the PPTA mostly relies on data above 800 MHz, NANOGrav uses 
narrow-band (25$-$50 MHz) low-frequency observations (430 MHz) in addition to the high-frequency 
observations (1.4 GHz and above) in their campaign. On the other hand, InPTA covers the low 
frequencies with wide-band receivers, where the dispersion is most prominent. This allows 
precision in-band DM estimates \citep[for example, see][]{Liu2014}. When combined with 
simultaneous higher-frequency observations, high precision DM estimates are possible. In this 
paper, we assess the usefulness of this combination for high precision DM measurements.

It is therefore of utmost importance to such experiments that the pulsar DMs be measured 
to high precision. As the DM delay scales with the observing frequency as 
$\Delta_{\text{DM}}\propto \text{DM}\; \nu^{-2}$, high precision DM measurements are 
possible at lower observing frequencies, although one must be mindful of certain caveats, 
such as the frequency dependence of the DM due to multi-path propagation through the IISM 
\citep{Cordes2016,Donner2019} and the effect of the variable scatter broadening of the pulse 
profiles observed at low radio frequencies while applying low-frequency DM measurements to 
correct ToAs measured at high frequencies \citep{levin2016}.

With the advent of a new generation of upgraded telescopes and their wide-band receivers, 
the attainable precision in DM measurements has greatly improved in recent years 
\citep[e.g.][]{Kaur2019,Tiburzi2019,DVT+20}. The Giant Metre-wave Radio Telescope 
\citep[GMRT;][]{Swarup1991} has recently gone through a major upgrade of its receivers and 
back-end instrumentation \citep[uGMRT;][]{Gupta2017,Reddy2017}, which has enabled an almost 
seamless frequency coverage from 120 to 1450 MHz. This improvement in the frequency coverage 
along with its capability of simultaneously observing a source at different frequency bands 
using multiple sub-arrays has greatly enhanced the precision with which the uGMRT can measure 
pulsar DMs. This enables the uGMRT to play an important role in eliminating low-frequency DM 
noise in PTA experiments.

In our technique, we use multiple profiles obtained across wide bandwidths for DM estimation. 
The DM obtained with this method will be insensitive to profile evolution over frequencies as 
the model template will be frequency-resolved in a similar manner. One important factor in 
getting the correct DM is the alignment of the sub-band profiles in the template. Small 
differences in the alignment can cause a systematic offset in the measured DM and will make 
a combination with other PTA datasets or application at higher frequencies difficult. In this 
paper, we discuss two different ways of aligning the wide-band profiles to measure in-band 
(\bandthree{} alone) and inter-band (\bandthree{} and \bandfive{} combined) DMs using data 
obtained by the uGMRT (details of the band definitions can be found in 
Section~\ref{sec:observations}).

The four pulsars for which we present our initial analysis are PSRs J1713+0747, J1909$-$3744, 
J1939+2134, and J2145$-$0750. PSR J2145$-$0750 has low solar elongations between December and 
February every year. This implies that the DM for this pulsar has an excess contribution from 
solar wind every year when it is close to the Sun 
\citep{Ujjwal2012,madison+19,Tiburzi2019,Tiburzi2020,ng12.5yr,DVT+20}. The DM can also be 
enhanced in case of a violent solar event, such as a coronal mass ejection (CME) or a 
CME-solar wind or CME-CME interaction, where the electron density in the line of sight can 
get enhanced. We report on such a  DM excess event observed on PSR J2145$-$0750 for the first 
time in our data.

The plan of the paper is as follows. The details of our observations are presented in 
Section \ref{sec:observations}. Our DM estimation methods are described in Section 
\ref{sec:analysis}, followed by results on individual pulsars in Section \ref{sec:results}. 
We compare the precision in DMs that we can achieve with the uGMRT with that of other PTAs and 
discuss our results in Section \ref{sec:conclusion}.

\section{Observation and data processing}
\label{sec:observations}

In this work, we use observations of four MSPs conducted between April 2018 and March 2019 as 
part of the InPTA campaign. PSRs J1713+0747, J1909$-$3744 and J1939+2134 were chosen for this 
study due to their significant long-term DM variations \citep{ng12.5yr,DVT+20}, while PSR 
J2145$-$0750 was chosen due to its high brightness for in-band analysis. Moreover, J1713+0747 
and J1909$-$3744 are two pulsars with the highest timing precision achieved in PTA experiments 
\citep{Verbiest2016,ng12.5yr}.

The pulsars were observed typically once every two weeks using the uGMRT in a multi-band phased 
array configuration. The 30 antennas of the uGMRT were split into three phased sub-arrays with 
the innermost 5 antennas used in \bandthree{} (400$-$500 MHz), 12 of the remaining outer 
antennas used in \bandfive{} (1360$-$1460 MHz), and another 8 used in \bandfour{} (650$-$750 
MHz). Each pulsar was observed in the three bands simultaneously at every epoch. The data in 
each band were acquired using a 100 MHz band-pass with 1024 sub-bands, where \bandthree{} and 
\bandfive{} data were coherently dedispersed using a real-time coherent dedispersion pipeline 
\citep{DeGupta2016} to the known DM of the pulsar. The coherently dedispersed data were sampled 
at 81.92 $\mu$s sampling time and recorded for further processing. In this work we only used 
the coherently dedispersed data obtained with \bandthree{} and \bandfive{} as the incoherently 
dedispersed \bandfour{} data were of much lower sensitivity for the in-band analysis described 
later. Further details on the available uGMRT configurations may be found in \cite{Gupta2017} 
and \citet{Reddy2017}.

\begin{figure*}
    \centering
    \includegraphics[scale=0.95]{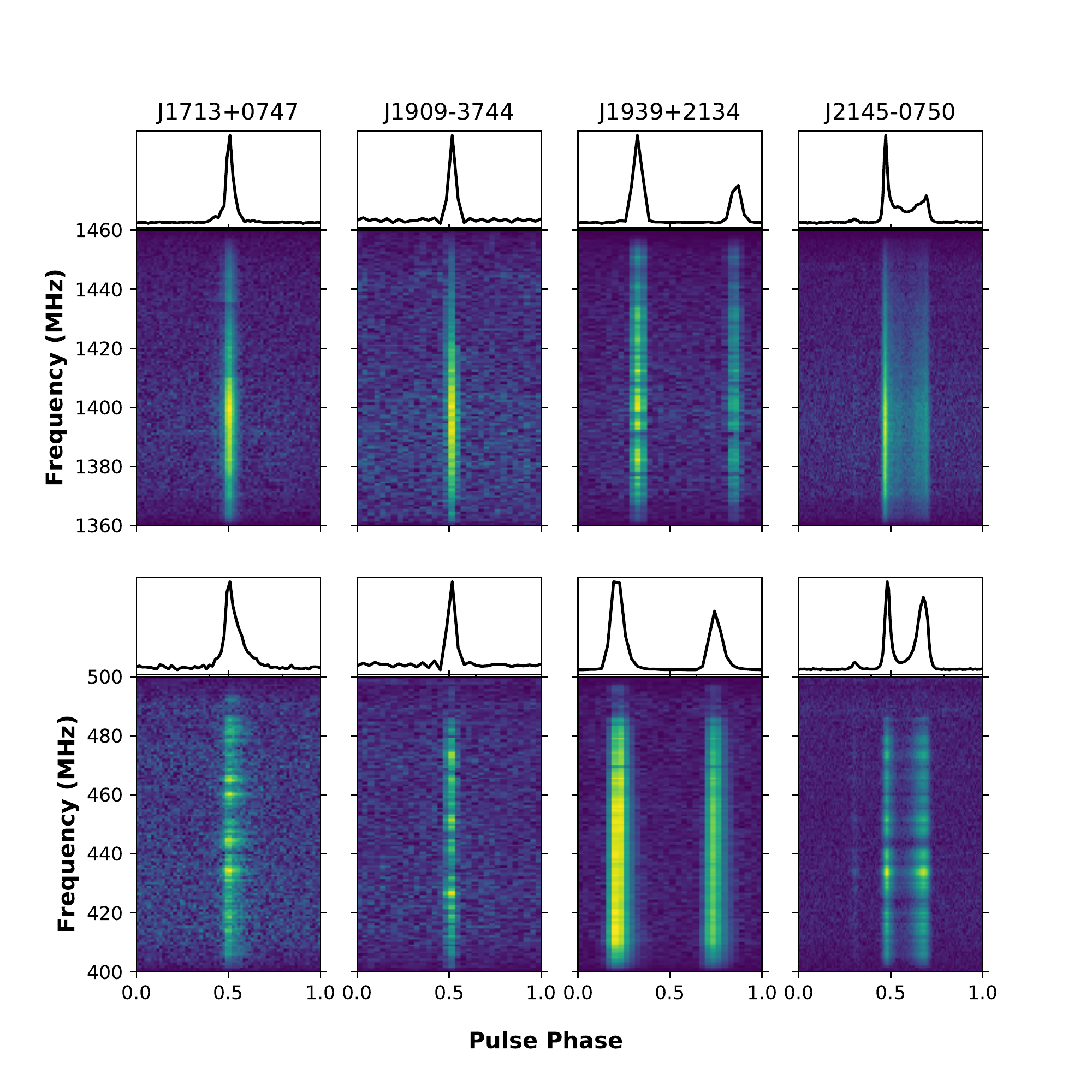}
    \caption{Collage of the frequency evolution seen in the pulse profiles of the four pulsars 
             presented in this work, along with their frequency averaged profiles. The top 
             panel shows the data from \bandfive{} and the bottom panel shows \bandthree{} data. 
             The data for the plot were obtained after adding high S/N observations from several 
             epochs together using the known ephemeris of each pulsar.}
    \label{fig:spectra_4psrs}
\end{figure*}

The timing mode data generated by the uGMRT were recorded using the GMRT Wide-band Backend 
\citep[GWB;][]{Reddy2017} in a raw data format. The timestamp of the start of the observation 
is written in a separate file, while all other relevant information about the observation, 
such as the observing frequency, bandwidth, sampling time, etc., can be accessed from a setup 
file created by the observer based on which the observation is carried out. The raw data are 
processed using the information from these files before it can be analysed by widely used pulsar 
software such as \psrchive{} \citep{Hotan2004}. We convert this raw data to the \texttt{Timer} 
format \citep{vanStraten2011} using a pipeline named 
\pinta{}\footnote{\url{https://github.com/abhisrkckl/pinta}} \citep{Susobhanan2020} developed 
for the InPTA campaign. \pinta{} performs radio frequency interference (RFI) mitigation using 
either \texttt{gptool}\footnote{\url{https://github.com/chowdhuryaditya/gptool}} 
\citep{ChoudharyGupta2020} or \rficlean{}\footnote{\url{https://github.com/ymaan4/rficlean}} 
\citep{Maan2020}, and folds the data using \dspsr{} \citep{vanStraten2011}, while supplying 
the required metadata (such as observing frequency and bandwidth) based on the observatory 
settings under which the observation was carried out. We supplied \dspsr{} with the pulsar 
models available from the IPTA Data Release 1 \citep{Verbiest2016} for folding. In the 
analysis presented in this work, we exclusively use \rficlean{} for RFI mitigation, which is 
designed to remove periodic RFI, such as the RFI caused by the 50~Hz power distribution grid 
as well as narrow-band and spiky RFI.

The details of the observations and the achieved profile signal to noise ratios (S/N) over the 
entire band are summarised in Table \ref{tab:observation_details}. Both in-band and inter-band 
estimates of the DM are presented in this work, which required reasonably high S/N ($> 30$) 
within individual sub-bands, and this was achieved on most epochs. A plot of the frequency 
evolution of the four pulsars used in this work and their integrated profiles in both the bands 
are shown in Figure~\ref{fig:spectra_4psrs}. Multiple high S/N observations were added together 
using the \psrchive{} tool \texttt{psradd} to obtain the data plotted in this figure.

\begin{table}
\caption{Summary of the observations used in this work. The table lists the duration of a 
typical observation, the median signal to noise ratio (S/N) of all the observations obtained 
in \bandthree{} and \bandfive{} using the \texttt{pdmp} program of \psrchive{} after removing 
the non-detections, and the total number of observations for each pulsar. The observations 
were carried out over a time period from April 2018 to March 2019.}
 \begin{tabular}{ c c c c c }
\hline\hline
PSR & Observation     & \multicolumn{2}{c}{Median S/N}  & No. of \\
    & duration (mins) & BAND3 & BAND5 & Epochs \\
\hline
J1713+0747   & 20 -- 25 &  40 &  80 & 17 \\
J1909$-$3744 & 20 -- 30 &  50 &  50 & 20 \\
J1939+2134   & 10 -- 15 & 270 & 110 & 20 \\
J2145$-$0750 & 10 -- 25 & 170 &  60 & 17 \\ \hline
\end{tabular}
\label{tab:observation_details}
\end{table}

\section{Data analysis}
\label{sec:analysis}

The data folded with \dspsr{} after removing the RFIs using \rficlean{} are directly used for 
estimating the DM. Due to the limited time span of the dataset ($\sim$1 year), it is not 
possible to obtain a reliable timing solution from these data. Hence, we used the latest 
parameter files published by the NANOGrav collaboration in their 12.5-year data release 
\citep{ng12.5yr} for estimating DM. The first requirement for obtaining a high precision DM 
measurement using wide-band data like ours is to obtain a frequency-resolved high S/N template
and aligning the sub-band profiles properly so that there is no residual DM delay in the 
template. If this correction is not done properly, the DMs estimated using such a template 
will be biased. We used two different methods to align the sub-band profiles in the template 
to check their effectiveness on the DM measurements as described in Section~\ref{sec:alignment}. 
We used these frequency-resolved templates to obtain ToAs and measure DM using \tempotwo{} 
\citep{Hobbs2006}. A Python-based script, \dmcalc{} was developed for this purpose using the 
\psrchive{} tools. We also implemented an outlier rejection algorithm for removing large 
outlier ToAs using Huber Regression \citep{Huber1964} following \citet{Tiburzi2019}. Details 
of our DM measurements are given in Section~\ref{sec:dmMethod}.

\begin{figure*}
    \centering
    \includegraphics[scale=0.8]{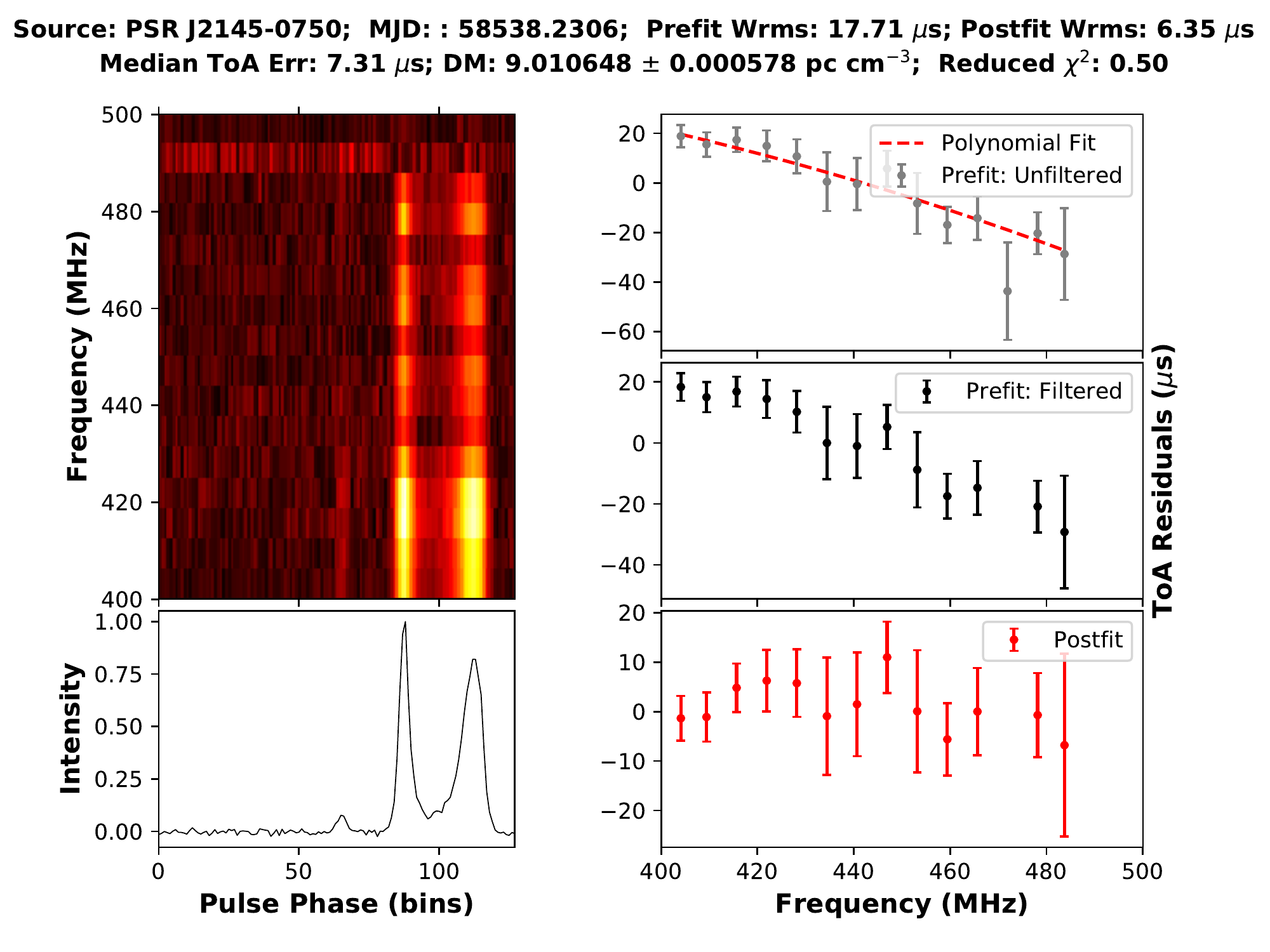}
    \caption{Sample analysis plot of \dmcalc{} using the observation of PSR J2145$-$0750 
             at \bandthree{} on 24 February 2019, when an excess DM was seen towards this 
             pulsar (see Section \ref{sec:results} for the discussion). Details of the fit 
             can be found at the top of the plot. \textbf{Right panel}: The top plot shows 
             the pre-fit residuals obtained from \tempotwo{} as grey circles and the Huber 
             Regression fit to it as a dashed line in red. The middle panel shows the 
             pre-fit ToAs after removing the outliers. The bottom panel shows the ToAs after 
             fitting for DM using \tempotwo{}. The details of the analysis method can be 
             found in Section~\ref{sec:dmMethod}. \textbf{Left panel}: The top panel shows 
             an image of the frequency spectra of the pulse profiles of a 25 min observation 
             after applying the DM correction and the bottom shows the frequency and time 
             averaged profile of the observation.
             }
    \label{fig:J2145_sample}
\end{figure*}

\subsection{Selection of the template and their alignment}
\label{sec:alignment}

In our first method (\methodone{}), we selected an epoch where the S/N (obtained using the 
\texttt{pdmp} program of \psrchive{}) of the observation is comparatively high at both bands 
(\bandthree{} and \bandfive{}). We estimated the DM at \bandthree{} using the \texttt{pdmp} 
program. Although the precision with which \texttt{pdmp} reports the DM is not very high, it 
is sufficient to align the sub-band profiles well in most cases. If the precision in the DM 
measurement reported by \texttt{pdmp} is worse than the change in DM from the ephemeris (with 
which the data are dedispersed), we did not update the DM (This is the case with PSR 
J1909$-$3744). The obtained DM is then used to dedisperse both \bandthree{} and \bandfive{} 
data. Smoothed templates were created from these files with the \texttt{psrsmooth} program in 
\psrchive{} using the wavelet smoothing algorithm \citep{DFG+13}. These smoothed templates 
were later used to estimate the DM.

It is possible that \methodone{} could bias the DM measurements as the alignment of the 
templates is performed  using the \texttt{pdmp} DM, which tries to maximise the S/N while 
obtaining the best DM. To circumvent this issue, we employed a different method (\methodtwo{}) 
for alignment, using an analytic template derived from the data. To do this, for every pulsar,  
we co-added the highest S/N observations in both  bands using \texttt{psradd} to create one 
final fiducial dataset. A frequency-averaged  as well as a  time-averaged profile was produced 
from this co-added data in  \bandthree{}. We then used the \psrchive{} tool \texttt{paas} to 
create an analytic template by fitting a mixture of  Gaussians to this  \bandthree{} profile. 
The noise-free analytic template created with this best fit was then used to estimate the DM 
of the aforementioned co-added dataset with \dmcalc{}. The sub-banded profiles in the high 
S/N data were then aligned using the DM obtained from \dmcalc{}, in both bands. We then used 
\texttt{psrsmooth} (similar to \methodone{}), to obtain a noise-free frequency-resolved 
template using the DM-corrected co-added data. The frequency-resolved templates produced 
using both these methods were then used to obtain the DM time series, as described in the 
next subsection.

The DMs obtained using \methodtwo{} have, in general, an order of magnitude better 
uncertainties than the ones obtained with \methodone{}. We also note that, in some cases, 
the actual DM value obtained using the two methods were slightly different. Additionally, 
it is possible for \methodtwo{} to give a biased DM for pulsars that show significant 
profile component evolution within the band as the initial alignment is obtained using a 
frequency-averaged profile.

\subsection{Measurement of the DM}
\label{sec:dmMethod}

To measure the DM, we used frequency-resolved templates prepared as explained in 
Section~\ref{sec:alignment}. This approach removes the need for fitting other 
frequency-dependent parameters while fitting for DM as the pulsar profile shape at a given 
frequency remains very much invariant (except for mode changes or scattering variations). 
The DMs reported in this paper are obtained using the \tempotwo{} package. We made use of 
the Python interface of \psrchive{} for obtaining the ToAs and also for removing the 
outliers. Most of the data processing was performed with this Python interface, except for 
obtaining the ToA residuals and for the fitting of the DM, which were performed using 
\tempotwo{}. The procedure for performing the outlier rejection we use here closely follows 
that by \citet{Tiburzi2019}. A Python-based tool named 
\dmcalc{}\footnote{\url{https://github.com/kkma89/dmcalc}} was developed for performing the 
above operations.

We used the latest parameter files published by \citet{ng12.5yr} for obtaining the DM. 
We removed the DM and the DMX parameters from the parameter files as this could otherwise 
bias the measured DM values. FD parameters were also removed as we perform frequency-resolved 
ToA estimation in this work. We also kept the electron density due to the solar wind (NE\_SW) 
as zero so as to remove bias. The DM in the parameter file was updated to the one that is 
obtained using either \methodone{} or \methodtwo{} for use in both the methods. The ToAs with 
the given frequency resolution for each pulsar were obtained at both bands by using the 
\texttt{ArrivalTime} class of \psrchive{} available with the Python interface. We used the 
classical Fourier phase shift estimation method \citep{Taylor92} implemented in \psrchive{} 
as \textit{PGS} for obtaining the ToAs. The ToAs thus obtained were then used to obtain 
frequency-resolved timing residuals using the \texttt{general2} plugin of \tempotwo{}. A fit 
of $\nu^{-2}$, where $\nu$ is the barycentric frequency of the ToAs was performed to these 
residuals using Huber Regression \citep{Huber1964}. A robust median absolute deviation (MAD) 
of the ToA residuals after removing the above fit from the residuals is calculated and the 
ToAs beyond three times the MAD value on both sides of the ToA residuals were removed. This 
outlier rejection method is effective in removing the large outliers that are otherwise present 
due to RFI or other issues in the data (for example, scintillation will make data of some 
channels almost unusable due to a very low S/N), which will corrupt the DMs obtained. These 
filtered ToAs were then used to fit for DM with \tempotwo{}.

An example analysis plot of PSR J2145$-$0750 is shown in Figure~\ref{fig:J2145_sample}. In 
this particular fit, we used a total of 16 sub-band profiles across the available 100~MHz 
bandwidth. The top two sub-bands were removed from the template as they were contaminated 
by RFI at most of the epochs. A total of 14 ToAs were obtained, one of which was rejected 
based on the outlier rejection criteria discussed above. A fit for DM was performed and the 
resulting ToAs after removing the DM trend can be found at the bottom-right panel of the 
figure. The pre-fit and post-fit weighted RMS can be found at the top of the panel, in addition 
to other parameters. As can be seen from the figure, the weighted RMS improved after fitting 
for the DM and its value is close to the median ToA uncertainty.

This process is performed at \bandthree{} and \bandfive{} separately as well as in a combined 
\bandthree{} + \bandfive{} mode to obtain DMs. In the combined \bandthree{} + \bandfive{} mode, 
the data, as well as the templates of both bands, were combined with the modified pulsar 
ephemeris from \citet{ng12.5yr} using the \texttt{FrequencyAppend() function} available in the 
\psrchive{} Python interface. The addition of the data without the requirement of having any 
jumps between the two bands is justified as these were observed simultaneously using the same 
receiver chain and processed identically in the uGMRT correlator, which synchronises all antenna 
data using a one pulse per second clock signal from the observatory’s active hydrogen maser. The 
relative delay was experimentally determined to be zero up to a precision of 5 ns using 
engineering tests \citep{Susobhanan2020}. The procedure was then repeated for all the 
observations to obtain the DM time series as shown in Figure~\ref{fig:DMtimeseries}. In the case 
of inter-band DM measurements, the data of both bands were aligned using the pulsar ephemeris 
before obtaining the ToAs and DM.

The four pulsars presented in this work have different frequency evolution of their parameters 
like flux density, profile shapes, scintle sizes and scatter broadening. To illustrate this, 
frequency-resolved profiles for all four pulsars are shown in Figure~\ref{fig:spectra_4psrs}. 
As a result, we had to obtain ToAs with different frequency resolution for each of them as 
described in Section~\ref{sec:results}.

\begin{table*}
\centering
\caption{Median values of ToA uncertainties and DM. The ToA uncertainty for each pulsar is 
         shown using the frequency and time averaged profiles with the analytic template 
         created in \methodtwo{} for both bands. The median DM and DM uncertainty obtained 
         from the DM time series of each pulsar using \methodone{} and \methodtwo{} are also 
         given.}
 \begin{tabular}{ c | c c | l l l | l l l }
\hline\hline
PSR & \multicolumn{2}{|c|}{$\sigma_{TOA}$ ($\mu$s)} & \multicolumn{3}{|c|}{DM [\methodone{}] (pc cm$^{-3})$} & \multicolumn{3}{|c}{DM [\methodtwo{}] (pc cm$^{-3})$} \\
    & BAND3 & BAND5 & BAND3 & BAND5 & Combined & BAND3 & BAND5 & Combined \\
\hline
J1713+0747   & 3.2 & 1.2 & 15.991(2)   & 15.99(2)  & 15.9918(4)  & 15.991(2)   & 15.99(1)  & 15.9900(2) \\
J1909$-$3744 & 2.1 & 2.1 & 10.389(2)   & 10.46(6)  & 10.3900(4)  & 10.390(3)   & 10.30(11) & 10.3878(5) \\
J1939+2134   & 0.9 & 0.4 & 71.01672(9) & 71.011(3) & 71.01661(3) & 71.02325(9) & 71.022(4) & 71.02267(5)\\
J2145$-$0750 & 1.7 & 3.7 & 8.995(1)    & 8.95(9)   & 8.9941(3)   & 9.0048(7)   & 9.00(8)   & 9.0051(3)  \\ \hline
\end{tabular}
\label{tab:results_details}
\end{table*}

\section{Results and discussions}
\label{sec:results}

\begin{figure*}
\centering
\begin{minipage}[b]{0.49\linewidth}
\includegraphics[width=\linewidth, angle=0.0]{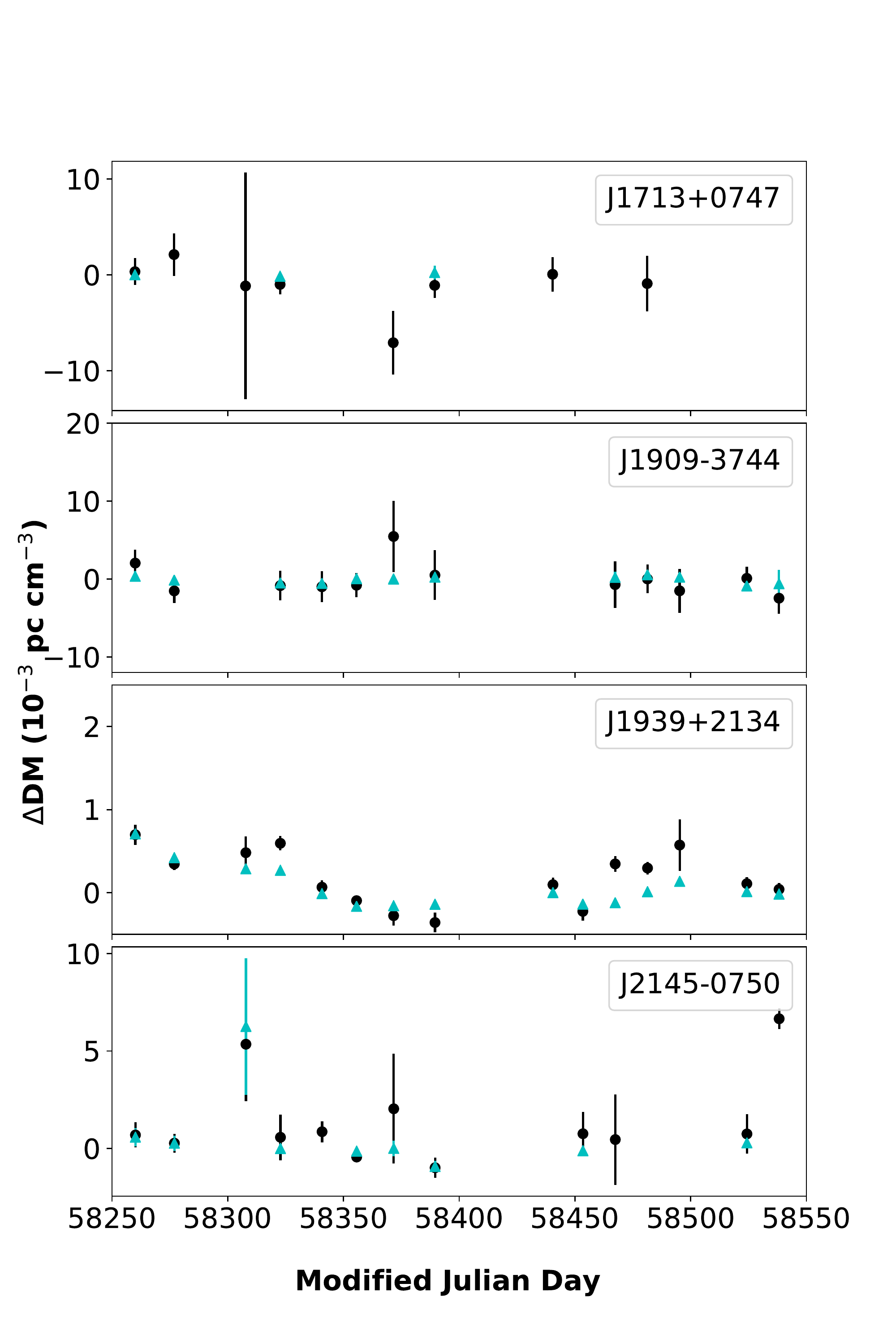}
\end{minipage}
\hfill
\begin{minipage}[b]{0.49\linewidth}
\includegraphics[width=\linewidth, angle=0]{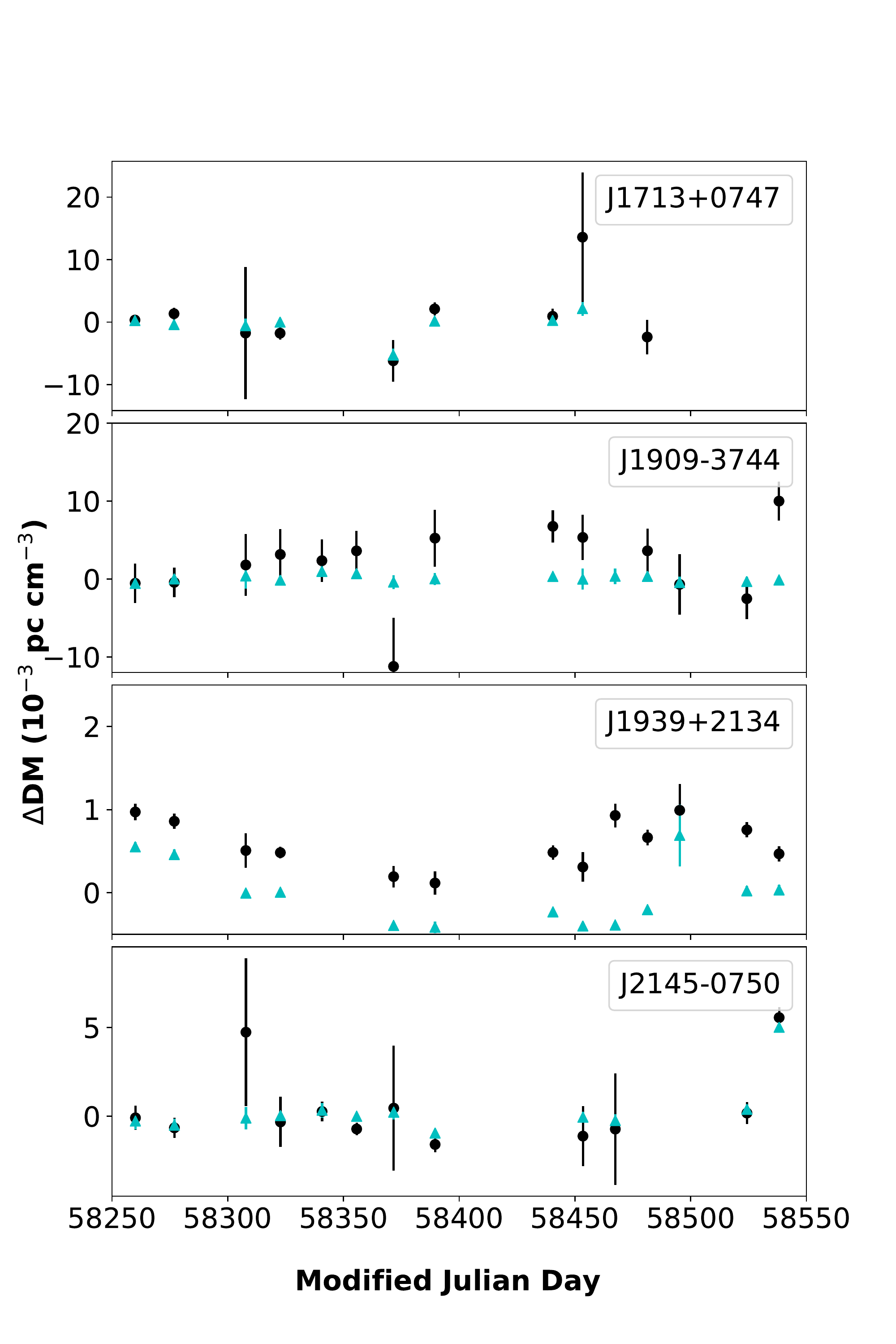}
\end{minipage}
\caption{DM time series after subtracting the median DM obtained from the \bandthree{} and 
         \bandfive{} combined data as shown in Table~\ref{tab:results_details} of the pulsars 
         presented in this work. The left panel shows the DM time series obtained by \methodone{}
         and Right panel shows the DM time series obtained by \methodtwo{}. Black filled circles 
         represent DM obtained from \bandthree{} and Cyan triangles indicate DM obtained by 
         combining bands 3 and 5. The median DM obtained from combining the two bands (refer to 
         Table~\ref{tab:results_details}) are subtracted from the DM values to produce this plot.
         The DMs obtained with only using \bandfive{} data are not shown in the plot as their 
         uncertainties are large.}
\label{fig:DMtimeseries}
\end{figure*}


The DM time series obtained using the methods described in Section~\ref{sec:results} for the 
four pulsars are shown in Figure~\ref{fig:DMtimeseries}. The left panel shows the DM measured 
using \methodone{} and the right panel shows DM obtained using \methodtwo{}. The median DM 
values and their uncertainties for the four pulsars are listed in 
Table~\ref{tab:results_details}. We have only reported the measurements for which a reduced 
$\chi^2 < 10$ is obtained with \tempotwo{}. Some epochs show reduced $\chi^2$ values worse than 
10, but looking at each of them individually showed that they were affected by heavy RFI. 
Although we put a higher limit on $\chi^2$ for getting the good measurements, most of them 
have reduced $\chi^2$ much less than 10 and close to 1. The median value of the DMs estimated 
using \methodone{} and \methodtwo{} differs slightly for PSRs J1713+0747 and J1909$-$3744, 
while there is a clear offset between the DMs for PSRs J1939+2134 and J2145$-$0750. The
possible cause of this difference is the underlying difference in the profile alignment 
methods used.

In the present work, we have used the data obtained by observing simultaneously at both 
\bandthree{} and \bandfive{} using a 100~MHz bandwidth. The fractional bandwidth at \bandfive{} 
is a factor of $\sim$4--5 smaller than those used by most of the other PTAs, which reduces the 
precision with which we can obtain DM at \bandfive{}. But a better fractional bandwidth at 
\bandthree{} enables us to get a good handle on the DM. The DM precision we can obtain in 
general by using this dataset with the \bandthree{} data alone is $\sim 10^{-3}$ and while 
combining the two bands it gets better by an order of magnitude to $\sim 10^{-4}$. In particular,
we achieve an order of magnitude better precision of 10$^{-4}$ with \bandthree{} and 10$^{-5}$ 
with \bandthree{} and \bandfive{} for PSR J1939+2134. Combining the two widely separated bands 
for measuring DM can create a bias due to the slightly different IISM the rays of these two 
bands pass through \citep{Cordes2016}.

Comparing our results obtained using the two methods described in this work to that of the 
recently published ones by \citet{ng12.5yr} and \citet{DVT+20} show interesting trends. For 
two pulsars, J1713+0747 and J2145$-$0750, we have data in both these datasets for comparison 
with ours. It should be noted that the data available from NANOGrav stops before our observations
began whereas the data from \citet{DVT+20} covers this gap as well as extends beyond our dataset.
For J1713+0747, we find our results from both  methods to be consistent with the results from
\citet{DVT+20}, whereas it shows a small increase in DM of $\sim2\times10^{-3}$ pc~cm$^{-3}$ 
from NANOGrav results. For J2145$-$0750, the DM from \methodone{} shows a difference of 
$\sim8\times10^{-3}$ pc~cm$^{-3}$, whereas the ones obtained with \methodtwo{} show consistency 
with the other two datasets. For J1909$-$3744 and J1939+2134, we only have DM measurements from 
NANOGrav to compare, although  the datasets do not overlap each other. For J1909$-$3744, the 
template DMs used for both methods are different due to their inherent differences in obtaining 
it. A difference of $\sim2\times10^{-3}$ pc~cm$^{-3}$ in the DM applied in the template caused 
the difference in the obtained DM using our two methods. Both of these measurements will have a 
small bias if the NANOGrav DM time series is extrapolated to cover our epochs. For J1939+2134, 
both alignment methods, \methodone{} as well as \methodtwo{}, could create a bias due to 
scattering. A completely different method taking care of the scattering evolution for each 
observation has to be used in such a case, which will be taken up in a follow-up work. In 
summary, both these alignment methods can be useful in getting DMs, but a systematic bias could 
be possible in either of the methods, which will be very much pulsar specific\footnote{We note 
that \citet{jones21} reported that the DMs obtained with the legacy GMRT system were offset 
compared to those obtained by NANOGrav, which they attributed to the frequency evolution of the 
pulse profile.}. Below we discuss in detail the results of each of the pulsars studied in this paper.

\subsection{PSR J1713+0747}

This is one of the most precisely timed pulsars in  PTA datasets. We did not detect this pulsar 
at some epochs. This could be due to the effect of diffractive scintillation affecting the 
luminosity at each epoch as the number of scintles available in the band becomes close to 1 
\citep{CordesLazio91,CordesChernoff97}. Assuming a scintillation bandwidth of 20 MHz and 
scintillation time of 45 min \citep{keith2013,levin2016}, the number of available scintles in 
the band at a given time will be ~1--3. In this low number regime, the pulse intensity is 
expected to be 100\% modulated \citep{CordesLazio91}, causing the S/N to be below our detection 
sensitivity at some epochs. This essentially reduced our DM precision at \bandfive{} and also 
made some of the observations essentially unusable for our analysis. We collapsed the data to 16 
channels at both  bands for obtaining the DMs. Both methods give similar DMs at both bands, but 
the combined estimate shows a small bias. The DMs used for aligning the templates using both the 
methods are slightly different, by $\sim$ 0.01 pc~cm$^{-3}$. From Figure~\ref{fig:DMtimeseries}, 
it can be seen that the DM measurements at some epochs are missing in the left panel. This is 
because the reduced $\chi^2$ of those fits are beyond the cutoff value and were removed from the 
plot. The average DM obtained in this work is consistent with that obtained by \citet{DVT+20} 
using LOFAR data, but is slightly higher than the DMs obtained by \citet{ng12.5yr} by about 
$2\times10^{-3}$  pc~cm$^{-3}$. This small bias from \citet{ng12.5yr} could be due to the 
frequency dependence of the DM (or scattering) as both \bandthree{} and LOFAR frequency bands 
are close to each other. The median ToA precision obtained at \bandfive{} is close to 1 $\mu$s.

\subsection{PSR J1909$-$3744}

Similar to the previous pulsar, this one is also a precisely timed pulsar with PTAs. Here also 
we collapsed the data to 16 channels at both bands for DM measurement. The average DM obtained 
using the two methods, after combining the two bands show a slight difference. This small bias, 
as in the previous case, could be due to the initial DM used for aligning the templates (they 
differ by $3\times10^{-3}$ pc~cm$^{-3}$). The pulse shape remains the same (without any major 
profile evolution) at both \bandthree{} and \bandfive{}. It is possible that we are unable to 
detect any small profile evolution due to the coarse sampling of the pulse phase. This prevented 
us from getting a better analytic profile for obtaining the DM with which the template was 
aligned. The DM time series reported in \citet{ng12.5yr} does not cover the epochs of our 
observations, but extrapolating their measurements to ours show a better alignment with the DMs 
obtained using \methodone{} and a small difference of $\sim 2\times10^{-3}$ pc~cm$^{-3}$ with that 
of \methodtwo{}, as evident from the difference in their average DMs. The ToA precision is similar 
in both bands.

\subsection{PSR J1939+2134}

This is one of the longest timed MSPs by all the PTAs \citep{kaspi94, Verbiest2016}. It shows 
timing noise in its ToA residuals and its timing data cannot be used for GW analysis without 
proper noise modelling. Since the pulsar is one of the brightest MSPs in our set, the precision 
in DM that can be achieved is quite high. Due to this, we used 128 channels at \bandthree{} and 
32 channels at \bandfive{} in the DM analysis. One limitation this pulsar has for using the 
\bandthree{} data for estimating DM is that it has very strong scatter broadening. Due to this 
reason, the initial DM obtained by the two different methods we used differ by about $\sim 
6\times 10^{-3}$ pc~cm$^{-3}$. This is exactly the difference between the average DMs reported 
in Table~\ref{tab:results_details} for the combined bands. There is a small difference of $\sim 
5\times10^{-4}$ pc~cm$^{-3}$ between the \bandthree{} DMs and the combined ones obtained using 
\methodtwo{}. This is probably due to the presence of scattering at \bandthree{}. The DM obtained
using both methods show differences even after taking these biases into account. This indicates 
that the scatter broadening present in the pulsar signal is also time-varying. A proper analysis 
of scatter broadening and simultaneous measurement of DM is required to disentangle the DM 
getting biased by the extra delay caused by scattering. This will be taken up in a future study. 
The DM time series obtained using \methodone{} follows the trend seen in \citet{ng12.5yr}, 
although the DMs reported here suffer from scattering bias. The ToA precision we obtain is the 
best for this pulsar in our sample, which is also indicative of the DM precision we could 
achieve.

\subsection{PSR J2145$-$0750}

This is one of the brightest pulsars in our sample. It shows a strong profile evolution across 
both of the bands. Moreover, this pulsar's line of sight passes close to the Sun at a solar 
elongation of $\sim$5 degrees. It has been reported previously that the DM shows an increase 
due to the increase in the heliospheric electron density (solar wind) as its line of sight 
approaches close to the Sun \citep{Ujjwal2012,ng12.5yr,DVT+20,Tiburzi2020}. Since this pulsar 
has several scintles in \bandthree{} data, we collapsed it to 16 channels to reduce the effect 
of scintillation. At \bandfive{}, we used 8 channels across the band. The median DMs obtained 
using the two methods shown in Table~\ref{tab:results_details} differ by about $1.1\times10^{-2}$
pc~cm$^{-3}$ for the combined bands. Even though the S/N of the data used for generating the 
template was good, the precision in DM using \methodone{} is worse than the difference quoted 
above. This is possibly due to change in relative amplitudes of profile components with 
frequency, which increases uncertainty while maximising S/N in \texttt{pdmp}. This is not an 
issue for \methodtwo{}, which obtained a precision in the fourth decimal place for the profile 
alignment using the analytic template. Although this creates a constant bias between the DM time 
series obtained using the two methods, the trend in it is not much affected as can be seen in 
Figure~\ref{fig:DMtimeseries}. The DMs obtained with \methodtwo{} show better alignment with 
the ones from \citet{ng12.5yr} and \citet{DVT+20}, while the ones obtained with \methodone{} 
have an offset. The median ToA precision we could obtain is about 4~$\mu$s at \bandfive{}.

\begin{figure}
\centering
\includegraphics[width=\linewidth]{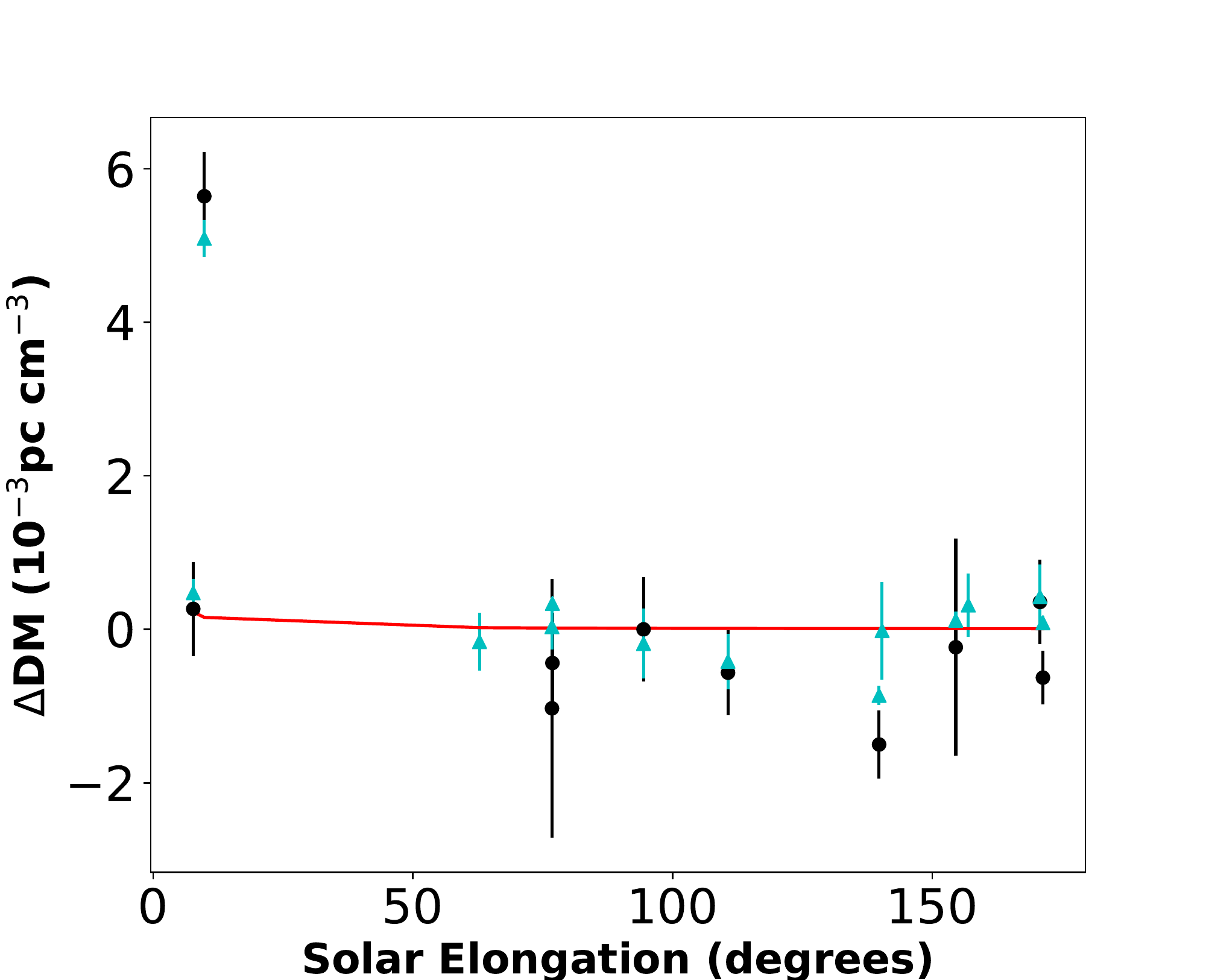}
\caption{DM time series of PSR J2145$-$0750 plotted as a function  of solar elongation after 
         subtracting the median DM value. The colour scheme is the same as in Figure 
         ~\ref{fig:DMtimeseries}. The red line shows the expected DM excess by the solar wind 
         as obtained from \tempotwo{}.}
    \label{fig:J2145solar}
\end{figure}

Since the line of sight to this pulsar passes close to the Sun (between January and March), 
we compare the observed DM time series as a function of solar elongation \citep[obtained from 
\tempotwo{} as \textit{solarangle};][]{You2007} as shown in Figure~\ref{fig:J2145solar}. The 
red curve in the figure shows the expected DM excess caused by the background solar wind as 
predicted by the model incorporated in \tempotwo{}. We have only two observations as the line 
of sight to the pulsar passed close to the Sun, respectively, at solar elongations $\sim$5 and 
$\sim$10 degrees. In Figure~\ref{fig:J2145solar}, it is seen that the DM measurement on 10 
February 2019 (MJD: 58524) at a solar elongation of $\sim$5 degrees shows nominal increase and 
it is consistent with the value expected from the model, whereas the other measurement at about 
10 degrees (i.e. a radial distance of $\sim$40 solar radii) away from the Sun on 24 February 
2019 (MJD: 58538) shows a DM excess of about an order of magnitude higher than the model. The 
\dmcalc{} fit for this excess DM observed is shown in  Figure~\ref{fig:J2145_sample}. To find 
the cause of this excess DM, we carefully examined the various solar datasets and solar wind 
measurements available during this epoch.

The examination of solar images from the Solar Dynamic Observatory \citep[SDO;][]{Sdo_paper} 
revealed the onsets of two eruptions, i.e. CMEs at $\sim$10 degrees west of the Sun's centre 
between 03 and 24 UT on 23 February 2019. The \textit{ahead} spacecraft of the Solar TErrestrial 
RElations Observatory \citep[STEREO-A;][]{kaiser2008} was located 99$^\circ$ east of the 
Sun-Earth line and it observed the above eruptions at about 20$^\circ$  behind the west limb 
of the Sun. Since these CMEs originated close to the disk centre and were relatively narrow, 
they did not fill and show their expansion outside the field of view of the occulting disk of 
the coronagraph at the near-Earth spacecraft. In addition to these CMEs, the SDO images showed 
the presence of a large coronal hole $\sim$30$^\circ$  wide, extending from the origin of the 
CME to the east nearly along the equatorial region of the Sun. The high-speed streams from the 
coronal hole were likely to interact with the low-speed solar wind as well as CMEs.

\begin{figure}
    \centering
    \includegraphics[width=\linewidth]{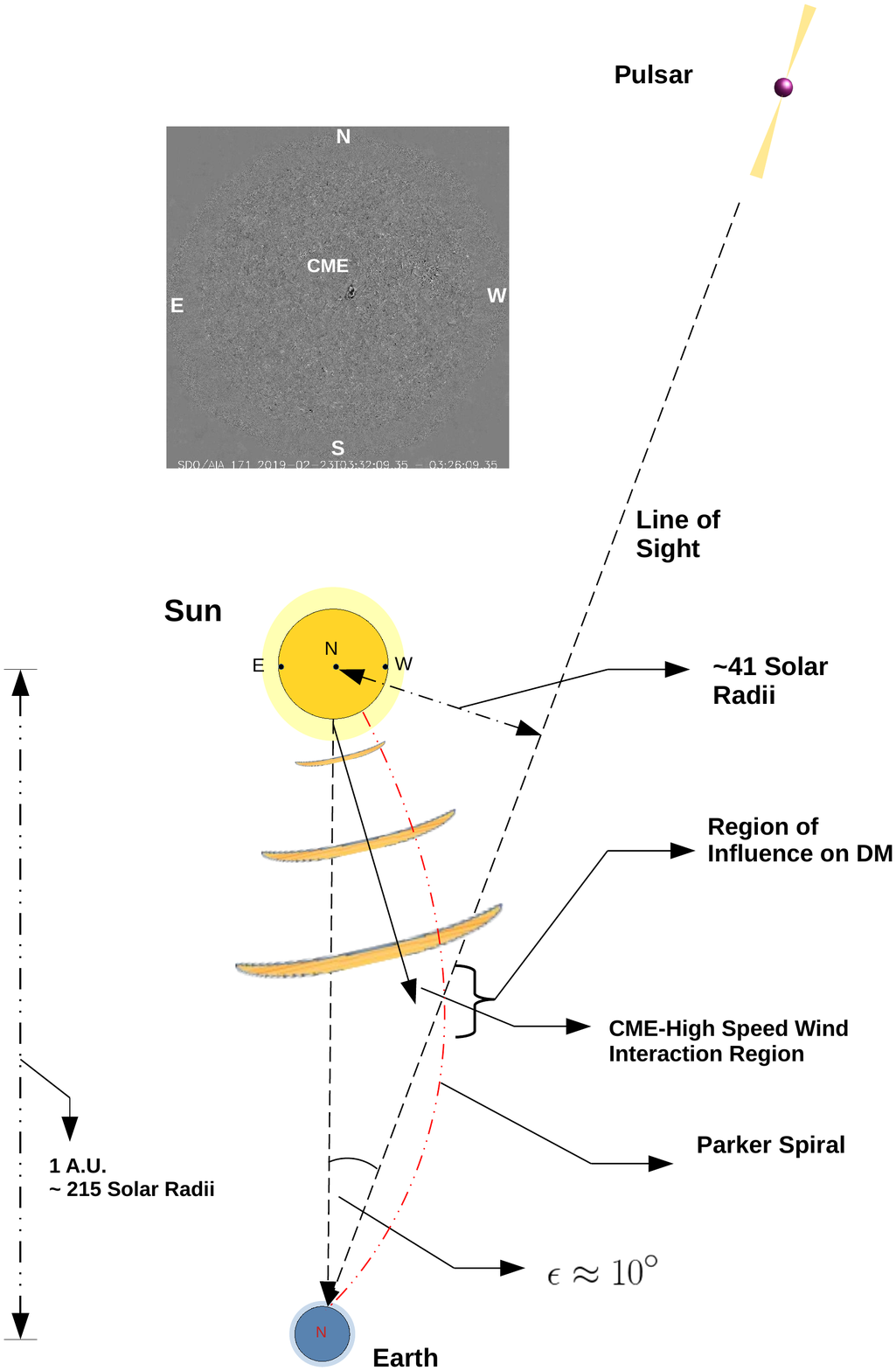}
    \caption{Sketch of the geometry of the line of sight to the pulsar with respect to the Sun 
             on 24 February 2019. The interaction region with excess electrons can be large in 
             size while crossing the line of sight. The inset image shown at the top is the 
             running difference EUV image of the Sun taken at 03:32 UT on 23 February 2019 by 
             the SDO AIA telescope at 171 {\AA}.
             }
    \label{fig:CME_cartoon}
\end{figure}

\begin{figure}
    \centering
    \includegraphics[width=\linewidth]{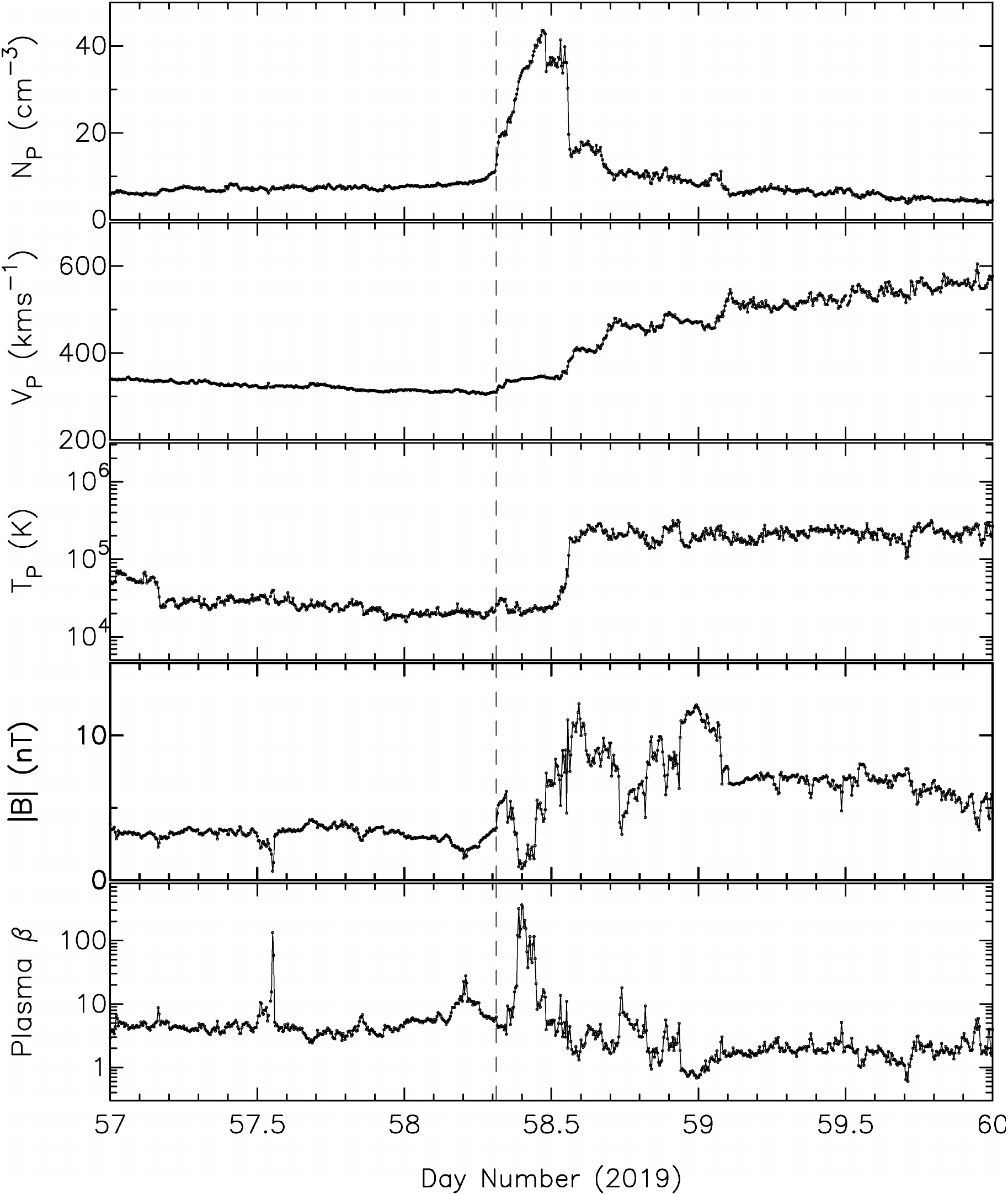}
    \caption{{\it In situ} measured 5-minute averaged OMNI data for a 3-day period, from 26 to 
             28 February 2019, during the passage of interaction region associated with the 
             high-speed solar wind streams and slow CME or ambient wind. From the top to bottom 
             the following data are plotted: solar wind proton density ($N_p$), velocity ($V_p$),
             temperature ($T_p$), magnitude of the interplanetary magnetic field ($|B|$) and 
             plasma beta ($\beta$). The vertical dashed line indicates the arrival of the 
             interplanetary shock associated with the interaction region. The time immediately 
             after the shock shows the intense interaction region, which is followed by the 
             clear signatures of the high-speed streams from the coronal hole. The data used in 
             this plot are obtained from the OMNI database at 
             \url{https://omniweb.gsfc.nasa.gov}.}
    \label{fig:omni_data}
\end{figure}

Figure~\ref{fig:CME_cartoon} shows the typical geometry of the line of sight to the pulsar with 
respect to the Sun, the possible propagation direction of CMEs, and slow solar wind along the 
Parker (Archimedean) spiral. The analysis of the interplanetary magnetic field and solar wind 
plasma from the OMNI datasets revealed an interplanetary shock at 07:35 UT on 27 February 
associated with the interaction between the slow- and high-speed solar wind streams. 
Figure~\ref{fig:omni_data} shows a 3-day period solar wind and interplanetary magnetic field 
measurements from 26 to 28 February 2019,  obtained from the OMNI 
database\footnote{\url{https://omniweb.gsfc.nasa.gov}}. From top to bottom, the figure shows the 
solar wind proton density, velocity, temperature, the magnitude of interplanetary magnetic field 
and plasma beta ($\beta$). The arrival of the shock is indicated by a vertical dotted line. 
The average ambient solar wind speed of $\sim$300 to 350 km/s, observed during the latter half 
of February 2019, suggests that the interaction by the high-speed streams of speed $\sim$600 to 
650 km s$^{-1}$, would have been formed and developed well ahead of its arrival at the Earth. 
The shock was followed by an intense interaction region, which was more than an order of 
magnitude denser than the ambient solar wind as well as about a half day wide in time. In the 
interaction region, the magnetic field exhibited large intensity fluctuations and the plasma 
beta, which is the ratio between the gas and magnetic pressures, also showed a large peak. The 
temperature, density and velocity measurements after the interaction region showed clear 
characteristics of the streams from the coronal hole. The backward projection of the interaction 
region suggests that the interaction would have crossed the pulsar line of sight on 24 February 
around 2 to 8 UT.

In the case of the ambient solar wind, the density decay with the distance from the Sun can be 
considered to be $R^{-2}$, typical for a spherically symmetric expansion of the solar wind, 
where $R$ is the heliocentric distance. However, when the high-density solar wind structures, 
such CMEs and/or high-speed stream interactions are involved, a radial density gradient of 
$R^{-2.5}$ or steeper has been observed \citep[e.g.][]{bird1994,Elliot2012}.

Assuming the $R^{-2}$ relation, it can be estimated that this interaction region had a density 
of $\sim1\times10^3$ cm$^{-3}$, taking 45~cm$^{-3}$ from the top panel of 
Figure~\ref{fig:omni_data} as the density at Earth (1 AU). The above density enhancement was 
also confirmed by the NOAA/NASA/USAF {\it Deep Space Climate Observatory} space 
mission\footnote{\url{https://spdf.gsfc.nasa.gov/pub/data/dscovr}}. This density region 
(assuming the same extent of the interaction region at 41 solar radii) will create an excess 
DM of $1\times10^{-3}$ pc~cm$^{-3}$. If we assume the steeper density gradient of $R^{-2.5}$, 
a DM excess of $3\times10^{-3}$ pc~cm$^{-3}$ can be obtained. Another point to be considered is 
that the eastern side of the interaction region likely crossed the Earth and it was possibly a 
little less dense than the nose of the interaction region, as indicated by the {\it in situ} 
measurements. Thus, the excessive DM observed probably corresponds to the density enhancement 
caused by the interactions between high-speed and low-speed solar wind and CMEs.

A similar CME event was reported on PSR B0950+08 by \citet{howard+16} and also possibly on PSR 
J1614$-$2230 by \citet{madison+19}. The solar wind stream interactions as well as stream-CME 
interactions are expected when the Sun is dominated by the mid-latitude and equatorial coronal 
holes. The vast sets of PTA and other pulsar observations available are likely to include many 
such events. A coordinated analysis of selected datasets would be of interest in understanding 
the effects of enhanced density structures of the solar wind on the DM variations as functions 
of solar offset and possibly also the phase of the solar cycle.

\section{Summary and conclusions}
\label{sec:conclusion}


In this paper, we compared the two possible methods for aligning the frequency-resolved pulsar 
profile templates and probed their effects on the resulting DM measurements. We used four InPTA 
pulsars observed by the uGMRT for a period of a year (two observing cycles at GMRT). These 
observations were done simultaneously at \bandthree{} and \bandfive{} of uGMRT with a 100~MHz 
bandwidth. For a uniform and systematic processing, we have developed a Python-based tool, 
\dmcalc{}, utilising the \psrchive{} Python interface and \tempotwo{} for estimating DM using 
the templates from the above two methods. We regularly obtained a DM precision of $\sim10^{-3}$ 
pc~cm$^{-3}$ at \bandthree{} and $\sim10^{-4}$ pc~cm$^{-3}$ when combining it with \bandfive{} 
data while using both of our template alignment methods.

We find that both the methods are useful for aligning the templates, but \methodtwo{} could 
show a constant bias if the pulsar has scatter broadening. For pulsars that have no detectable 
scatter broadening, the DMs obtained by \methodone{} show a consistent bias from that obtained 
with \methodtwo{}. This is essentially due to the use of two different methods for template 
alignment. \methodone{} uses an algorithm that aligns the multi-channel data by maximising the 
S/N while \methodtwo{} uses an analytic model derived from the frequency averaged profile of 
the data.

We have compared the DMs obtained by these two methods with the other recently published results 
\citep{ng12.5yr,DVT+20}. Our DM measurements for PSRs J1713+0747 and J2145$-$0750 using 
\methodtwo{} compare very well with theirs, while those from \methodone{} show a bias. In the 
cases of PSRs J1909$-$3744 and J1939+2134, our data have no overlap with either of the published 
datasets. Nevertheless, we see a continuing trend of the NANOGrav DM time series for J1939+2134 
using \methodone{}, while \methodtwo{} shows a clear constant offset. An improved method that 
takes care of the scattering bias while estimating the DM will be able to remove any bias created
by scattering. For J1909$-$3744, we expect to see a small offset from that of the NANOGrav data 
using both methods, although the DMs obtained with \methodone{} may have smaller bias than the 
other.

In addition to the DM offset introduced due to different template alignment methods, there could 
be other contributing factors for the DM offset that we have not been accounted for in this work.
One of the main causes for the offset between our DMs and the ones in the literature 
\citep{ng12.5yr,DVT+20} is due to scattering (at least for PSR J1939+2134). Scatter-broadening 
introduces an extra delay in the ToAs and it scales as $\nu^{-4}$. Furthermore, if the delay 
varies over time, it could corrupt the trend in the DM time series of the pulsar. Currently 
neither of our methods take that into account and this will be taken up in a future work. The DM 
obtained from combining the data from the two widely separated bands could also get affected by 
the profile evolution. Although we do not see such an offset in our data yet, given the coarse 
phase resolution, it is possible that we may have to take that into account when the phase 
resolution improves in future datasets. The DM obtained from combining the data from the two 
widely separated bands could also be affected by the profile evolution. Given the coarse phase 
resolution in our data, we currently do not see such an offset. However, it will have to be 
accounted for in future data with finer phase resolution. It was suggested by \citet{Cordes2016} 
that the DM is frequency dependent. This means that a possible offset could be introduced when 
these widely separated frequency bands are combined. Although we do not see any evidence for 
this with the current dataset, it could become evident with better datasets in the future.

Our data are in total intensity mode and hence there is no polarisation information available. 
This could affect the alignment of the combined \bandthree{} + \bandfive{} data. But our coarse 
pulse phase resolution and lack of full Stokes data makes it difficult to either confirm or rule 
it out.

We could obtain a ToA precision of $\sim1\mu$s or better for all pulsars in both the bands, 
which is highly encouraging. We see the effect of scattering on DM measurements of J1939+2134. 
In a follow-up study, we plan to disentangle the two effects to obtain better DM estimates.

We infer that the DM measurement at MJD 58538 of PSR J2145$-$0750 with a solar elongation of 
$\sim 10^{\circ}$ was enhanced by an interaction region formed by a CME and high-speed solar 
wind from a coronal hole close to the origin of the CME. Similar events can be of interest to 
both the pulsar and the solar wind community and our results show that such studies can be 
pursued using high precision data from the uGMRT.

The present observations used a 20 -- 25 mins scan for each pulsar. A much better precision on 
both ToAs and hence DM can be achieved by using longer integrations and wider bandwidths of the 
uGMRT. We started doing observations using a bandwidth of 200~MHz at both \bandthree{} and 
\bandfive{} simultaneously and also increased the observation duration in addition to increasing 
the number of antennas at each band (by skipping \bandfour{}, and utilising the antennas at the 
other two bands). A factor of three improvement in the precision of DM is expected at 
\bandthree{} in general with this increased bandwidth as compared to current results. Initial 
results show vast improvement in the S/N of the pulsars. The data from these observations are 
under various stages of processing and will be reported elsewhere.

Following the encouraging results from the work presented here, we plan to apply these 
techniques to our full sample of pulsars observed during the last four years. Additionally, 
efforts are being pursued for developing other methods to make the DM measurements even more 
precise and reliable, and therefore employable for the ongoing gravitational wave analysis by 
the various PTAs.

\section*{Data availability}
The data used in this paper will be made available on reasonable request. The SDO 171~{\AA} 
images used for the solar wind analysis can be found at 
\url{https://cdaw.gsfc.nasa.gov/movie/make_javamovie.php?date=20190223&img1=sdo_a304&img2=lasc2rdf}.

\section*{Acknowledgements}
MAK is thankful to Caterina Tiburzi and Joris Verbiest for their valuable inputs and useful 
discussions at various stages of this work.  We are thankful to the anonymous referee for many 
constructive comments on the manuscript. AS, AG, BCJ, LD, and YG acknowledge the support of the 
Department of Atomic Energy, Government of India, under project Identification \# RTI 4002. BCJ, 
YG and AB acknowledge support from the Department of Atomic Energy, Government of India, under 
project \# 12-R\&D-TFR-5.02-0700. AC acknowledge support from the Women's Scientist scheme 
(WOS-A), Department of Science \& Technology, India. MPS acknowledges funding from the European 
Research Council (ERC) under the European Union's Horizon 2020 research and innovation programme 
(grant agreement No. 694745). NDB acknowledge support from the Department of Science \& 
Technology, Government of India, grant SR/WOS-A/PM-1031/2014. AB acknowledges the support from 
the UK Science and Technology Facilities Council (STFC). Pulsar research at Jodrell Bank Centre 
for Astrophysics and Jodrell Bank Observatory is supported by a consolidated grant from STFC. We 
thank the staff of the GMRT who made our observations possible. GMRT is run by the National 
Centre for Radio Astrophysics of the Tata Institute of Fundamental Research. The open data policy
of STEREO and SDO teams is acknowledged. The solar wind and interplanetary datasets have been 
obtained from the OMNI database. 

\balance

\bibliographystyle{aa}
\bibliography{InPTAhighprecuGMRTDM}

\begin{thebibliography}{52}
\expandafter\ifx\csname natexlab\endcsname\relax\def\natexlab#1{#1}\fi

\bibitem[{{Ahuja} {et~al.}(2005){Ahuja}, {Gupta}, {Mitra}, \&
  {Kembhavi}}]{Ahuja2005}
{Ahuja}, A.~L., {Gupta}, Y., {Mitra}, D., \& {Kembhavi}, A.~K. 2005, \mnras,
  357, 1013

\bibitem[{{Ahuja} {et~al.}(2007){Ahuja}, {Mitra}, \& {Gupta}}]{Ahuja2007}
{Ahuja}, A.~L., {Mitra}, D., \& {Gupta}, Y. 2007, \mnras, 377, 677

\bibitem[{{Alam} {et~al.}(2021){Alam}, {Arzoumanian}, {Baker}, {Blumer},
  {Bohler}, {Brazier}, {Brook}, {Burke-Spolaor}, {Caballero}, {Camuccio},
  {Chamberlain}, {Chatterjee}, {Cordes}, {Cornish}, {Crawford}, {Cromartie},
  {Decesar}, {Demorest}, {Dolch}, {Ellis}, {Ferdman}, {Ferrara}, {Fiore},
  {Fonseca}, {Garcia}, {Garver-Daniels}, {Gentile}, {Good}, {Gusdorff},
  {Halmrast}, {Hazboun}, {Islo}, {Jennings}, {Jessup}, {Jones}, {Kaiser},
  {Kaplan}, {Kelley}, {Key}, {Lam}, {Lazio}, {Lorimer}, {Luo}, {Lynch},
  {Madison}, {Maraccini}, {McLaughlin}, {Mingarelli}, {Ng}, {Nguyen}, {Nice},
  {Pennucci}, {Pol}, {Ramette}, {Ransom}, {Ray}, {Shapiro-Albert}, {Siemens},
  {Simon}, {Spiewak}, {Stairs}, {Stinebring}, {Stovall}, {Swiggum}, {Taylor},
  {Tripepi}, {Vallisneri}, {Vigeland}, {Witt}, {Zhu}, \& {Nanograv
  Collaboration}}]{ng12.5yr}
{Alam}, M.~F., {Arzoumanian}, Z., {Baker}, P.~T., {et~al.} 2021, \apjs, 252, 5

\bibitem[{{Arzoumanian} {et~al.}(2020){Arzoumanian}, {Baker}, {Blumer},
  {B{\'e}csy}, {Brazier}, {Brook}, {Burke-Spolaor}, {Chatterjee}, {Chen},
  {Cordes}, {Cornish}, {Crawford}, {Cromartie}, {Decesar}, {Demorest}, {Dolch},
  {Ellis}, {Ferrara}, {Fiore}, {Fonseca}, {Garver-Daniels}, {Gentile}, {Good},
  {Hazboun}, {Holgado}, {Islo}, {Jennings}, {Jones}, {Kaiser}, {Kaplan},
  {Kelley}, {Key}, {Laal}, {Lam}, {Lazio}, {Lorimer}, {Luo}, {Lynch},
  {Madison}, {McLaughlin}, {Mingarelli}, {Ng}, {Nice}, {Pennucci}, {Pol},
  {Ransom}, {Ray}, {Shapiro-Albert}, {Siemens}, {Simon}, {Spiewak}, {Stairs},
  {Stinebring}, {Stovall}, {Sun}, {Swiggum}, {Taylor}, {Turner}, {Vallisneri},
  {Vigeland}, {Witt}, \& {Nanograv Collaboration}}]{Arzoumanian2020}
{Arzoumanian}, Z., {Baker}, P.~T., {Blumer}, H., {et~al.} 2020, \apjl, 905, L34

\bibitem[{{Backer}(1996)}]{Backer1996}
{Backer}, D.~C. 1996, in Compact Stars in Binaries, ed. J.~{van Paradijs},
  E.~P.~J. {van den Heuvel}, \& E.~{Kuulkers}, Vol. 165, 197

\bibitem[{{Bird} {et~al.}(1994){Bird}, {Volland}, {Paetzold}, {Edenhofer},
  {Asmar}, \& {Brenkle}}]{bird1994}
{Bird}, M.~K., {Volland}, H., {Paetzold}, M., {et~al.} 1994, \apj, 426, 373

\bibitem[{Burke-Spolaor {et~al.}(2019)Burke-Spolaor, Taylor, Charisi, Dolch,
  Hazboun, Holgado, Kelley, Lazio, Madison, McMann, Mingarelli, Rasskazov,
  Siemens, Simon, \& Smith}]{Burke-Spolaor2019}
Burke-Spolaor, S., Taylor, S.~R., Charisi, M., {et~al.} 2019, Astronomy and
  Astrophysics Review, 27, 5

\bibitem[{{Chowdhury} \& {Gupta}(2021)}]{ChoudharyGupta2020}
{Chowdhury}, A. \& {Gupta}, Y. 2021, In preparation

\bibitem[{{Cordes} \& {Chernoff}(1997)}]{CordesChernoff97}
{Cordes}, J.~M. \& {Chernoff}, D.~F. 1997, \apj, 482, 971

\bibitem[{{Cordes} \& {Lazio}(1991)}]{CordesLazio91}
{Cordes}, J.~M. \& {Lazio}, T.~J. 1991, \apj, 376, 123

\bibitem[{{Cordes} {et~al.}(2016){Cordes}, {Shannon}, \&
  {Stinebring}}]{Cordes2016}
{Cordes}, J.~M., {Shannon}, R.~M., \& {Stinebring}, D.~R. 2016, \apj, 817, 16

\bibitem[{De \& Gupta(2016)}]{DeGupta2016}
De, K. \& Gupta, Y. 2016, Experimental Astronomy, 41, 67

\bibitem[{{Demorest} {et~al.}(2013){Demorest}, {Ferdman}, {Gonzalez}, {Nice},
  {Ransom}, {Stairs}, {Arzoumanian}, {Brazier}, {Burke-Spolaor}, {Chamberlin},
  {Cordes}, {Ellis}, {Finn}, {Freire}, {Giampanis}, {Jenet}, {Kaspi}, {Lazio},
  {Lommen}, {McLaughlin}, {Palliyaguru}, {Perrodin}, {Shannon}, {Siemens},
  {Stinebring}, {Swiggum}, \& {Zhu}}]{DFG+13}
{Demorest}, P.~B., {Ferdman}, R.~D., {Gonzalez}, M.~E., {et~al.} 2013, \apj,
  762, 94

\bibitem[{Desvignes {et~al.}(2016)Desvignes, Caballero, Lentati, Verbiest,
  Champion, Stappers, Janssen, Lazarus, Os{\l}owski, Babak, Bassa, Brem,
  Burgay, Cognard, Gair, Graikou, Guillemot, Hessels, Jessner, Jordan,
  Karuppusamy, Kramer, Lassus, Lazaridis, Lee, Liu, Lyne, McKee, Mingarelli,
  Perrodin, Petiteau, Possenti, Purver, Rosado, Sanidas, Sesana, Shaifullah,
  Smits, Taylor, Theureau, Tiburzi, Van~Haasteren, \& Vecchio}]{Desvignes2016}
Desvignes, G., Caballero, R.~N., Lentati, L., {et~al.} 2016, Monthly Notices of
  the Royal Astronomical Society, 458, 3341

\bibitem[{{Donner} {et~al.}(2020){Donner}, {Verbiest}, {Tiburzi},
  {Os{\l}owski}, {K{\"u}nsem{\"o}ller}, {Bak Nielsen}, {Grie{\ss}meier},
  {Serylak}, {Kramer}, {Anderson}, {Wucknitz}, {Keane}, {Kondratiev}, {Sobey},
  {McKee}, {Bilous}, {Breton}, {Br{\"u}ggen}, {Ciardi}, {Hoeft}, {van Leeuwen},
  \& {Vocks}}]{DVT+20}
{Donner}, J.~Y., {Verbiest}, J.~P.~W., {Tiburzi}, C., {et~al.} 2020, \aap, 644,
  A153

\bibitem[{{Donner} {et~al.}(2019){Donner}, {Verbiest}, {Tiburzi},
  {Os{\l}owski}, {Michilli}, {Serylak}, {Anderson}, {Horneffer}, {Kramer},
  {Grie{\ss}meier}, {K{\"u}nsem{\"o}ller}, {Hessels}, {Hoeft}, \&
  {Miskolczi}}]{Donner2019}
{Donner}, J.~Y., {Verbiest}, J.~P.~W., {Tiburzi}, C., {et~al.} 2019, \aap, 624,
  A22

\bibitem[{Edwards {et~al.}(2006)Edwards, Hobbs, \& Manchester}]{Edwards2006}
Edwards, R.~T., Hobbs, G.~B., \& Manchester, R.~N. 2006, Monthly Notices of the
  Royal Astronomical Society, 372, 1549

\bibitem[{{Elliott} {et~al.}(2012){Elliott}, {Henney}, {McComas}, {Smith}, \&
  {Vasquez}}]{Elliot2012}
{Elliott}, H.~A., {Henney}, C.~J., {McComas}, D.~J., {Smith}, C.~W., \&
  {Vasquez}, B.~J. 2012, Journal of Geophysical Research (Space Physics), 117,
  A09102

\bibitem[{{Foster} \& {Backer}(1990)}]{FosterBacker1990}
{Foster}, R.~S. \& {Backer}, D.~C. 1990, \apj, 361, 300

\bibitem[{{Gupta} {et~al.}(2017){Gupta}, {Ajithkumar}, {Kale}, {Nayak},
  {Sabhapathy}, {Sureshkumar}, {Swami}, {Chengalur}, {Ghosh},
  {Ishwara-Chandra}, {Joshi}, {Kanekar}, {Lal}, \& {Roy}}]{Gupta2017}
{Gupta}, Y., {Ajithkumar}, B., {Kale}, H.~S., {et~al.} 2017, Current Science,
  113, 707

\bibitem[{Hobbs(2013)}]{Hobbs2013}
Hobbs, G. 2013, Classical and Quantum Gravity, 30, 224007

\bibitem[{Hobbs {et~al.}(2019)Hobbs, Guo, Caballero, Coles, Lee, Manchester,
  Reardon, Matsakis, Tong, Arzoumanian, Bailes, Bassa, Bhat, Brazier,
  Burke-Spolaor, Champion, Chatterjee, Cognard, Dai, Desvignes, Dolch, Ferdman,
  Graikou, Guillemot, Janssen, Keith, Kerr, Kramer, Lam, Liu, Lyne, Lazio,
  Lynch, McKee, McLaughlin, Mingarelli, Nice, Os{\l}owski, Pennucci, Perera,
  Perrodin, Possenti, Russell, Sanidas, Sesana, Shaifullah, Shannon, Simon,
  Spiewak, Stairs, Stappers, Swiggum, Taylor, Theureau, Toomey, van Haasteren,
  Wang, Wang, \& Zhu}]{Hobbs2020_IPTA_clock}
Hobbs, G., Guo, L., Caballero, R.~N., {et~al.} 2019, Monthly Notices of the
  Royal Astronomical Society, 491, 5951

\bibitem[{Hobbs {et~al.}(2006)Hobbs, Edwards, \& Manchester}]{Hobbs2006}
Hobbs, G.~B., Edwards, R.~T., \& Manchester, R.~N. 2006, Monthly Notices of the
  Royal Astronomical Society, 369, 655

\bibitem[{Hotan {et~al.}(2004)Hotan, van Straten, \& Manchester}]{Hotan2004}
Hotan, A.~W., van Straten, W., \& Manchester, R.~N. 2004, Publications of the
  Astronomical Society of Australia, 21, 302–309

\bibitem[{{Howard} {et~al.}(2016){Howard}, {Stovall}, {Dowell}, {Taylor}, \&
  {White}}]{howard+16}
{Howard}, T.~A., {Stovall}, K., {Dowell}, J., {Taylor}, G.~B., \& {White},
  S.~M. 2016, \apj, 831, 208

\bibitem[{Huber(1964)}]{Huber1964}
Huber, P.~J. 1964, Ann. Math. Statist., 35, 73

\bibitem[{{Jones} {et~al.}(2020){Jones}, {McLaughlin}, {Roy}, {Lam}, {Cordes},
  {Kaplan}, {Bhattacharyya}, \& {Levin}}]{jones21}
{Jones}, M.~L., {McLaughlin}, M.~A., {Roy}, J., {et~al.} 2020, arXiv e-prints,
  arXiv:2009.08409

\bibitem[{{Joshi} {et~al.}(2018){Joshi}, {Arumugasamy}, {Bagchi},
  {Bandyopadhyay}, {Basu}, {Dhanda Batra}, {Bethapudi}, {Choudhary}, {De},
  {Dey}, {Gopakumar}, {Gupta}, {Krishnakumar}, {Maan}, {Manoharan}, {Naidu},
  {Nandi}, {Pathak}, {Surnis}, \& {Susobhanan}}]{Joshi2018}
{Joshi}, B.~C., {Arumugasamy}, P., {Bagchi}, M., {et~al.} 2018, Journal of
  Astrophysics and Astronomy, 39, 51

\bibitem[{{Kaiser} {et~al.}(2008){Kaiser}, {Kucera}, {Davila}, {St. Cyr},
  {Guhathakurta}, \& {Christian}}]{kaiser2008}
{Kaiser}, M.~L., {Kucera}, T.~A., {Davila}, J.~M., {et~al.} 2008, \ssr, 136, 5

\bibitem[{{Kaspi} {et~al.}(1994){Kaspi}, {Taylor}, \& {Ryba}}]{kaspi94}
{Kaspi}, V.~M., {Taylor}, J.~H., \& {Ryba}, M.~F. 1994, \apj, 428, 713

\bibitem[{{Kaur} {et~al.}(2019){Kaur}, {Bhat}, {Tremblay}, {Shannon},
  {McSweeney}, {Ord}, {Beardsley}, {Crosse}, {Emrich}, {Franzen}, {Horsley},
  {Johnston-Hollitt}, {Kaplan}, {Kenney}, {Morales}, {Pallot}, {Steele},
  {Tingay}, {Trott}, {Walker}, {Wayth}, {Williams}, \& {Wu}}]{Kaur2019}
{Kaur}, D., {Bhat}, N.~D.~R., {Tremblay}, S.~E., {et~al.} 2019, \apj, 882, 133

\bibitem[{{Keith} {et~al.}(2013){Keith}, {Coles}, {Shannon}, {Hobbs},
  {Manchester}, {Bailes}, {Bhat}, {Burke-Spolaor}, {Champion}, {Chaudhary},
  {Hotan}, {Khoo}, {Kocz}, {Os{\l}owski}, {Ravi}, {Reynolds}, {Sarkissian},
  {van Straten}, \& {Yardley}}]{keith2013}
{Keith}, M.~J., {Coles}, W., {Shannon}, R.~M., {et~al.} 2013, \mnras, 429, 2161

\bibitem[{Kerr {et~al.}(2020)Kerr, Reardon, Hobbs, Shannon, Manchester, Dai,
  Russell, Zhang, van Straten, Os{\l}owski, Parthasarathy, Spiewak, Bailes,
  Bhat, Cameron, Coles, Dempsey, Deng, Goncharov, Kaczmarek, Keith, Lasky,
  Lower, Preisig, Sarkissian, Toomey, Wang, Wang, Zhang, \& Zhu}]{Kerr2020}
Kerr, M., Reardon, D.~J., Hobbs, G., {et~al.} 2020, Publications of the
  Astronomical Society of Australia, 37, e020

\bibitem[{Kramer \& Champion(2013)}]{KramerChampion2013}
Kramer, M. \& Champion, D.~J. 2013, Classical and Quantum Gravity, 30, 224009

\bibitem[{{Kumar} {et~al.}(2013){Kumar}, {Gupta}, {van Straten}, {Os{\l}owski},
  {Roy}, {Bhat}, {Bailes}, \& {Keith}}]{Ujjwal2012}
{Kumar}, U., {Gupta}, Y., {van Straten}, W., {et~al.} 2013, in Neutron Stars
  and Pulsars: Challenges and Opportunities after 80 years, ed. J.~{van
  Leeuwen}, Vol. 291, 432--434

\bibitem[{{Levin} {et~al.}(2016){Levin}, {McLaughlin}, {Jones}, {Cordes},
  {Stinebring}, {Chatterjee}, {Dolch}, {Lam}, {Lazio}, {Palliyaguru},
  {Arzoumanian}, {Crowter}, {Demorest}, {Ellis}, {Ferdman}, {Fonseca},
  {Gonzalez}, {Jones}, {Nice}, {Pennucci}, {Ransom}, {Stairs}, {Stovall},
  {Swiggum}, \& {Zhu}}]{levin2016}
{Levin}, L., {McLaughlin}, M.~A., {Jones}, G., {et~al.} 2016, \apj, 818, 166

\bibitem[{{Liu} {et~al.}(2014){Liu}, {Desvignes}, {Cognard}, {Stappers},
  {Verbiest}, {Lee}, {Champion}, {Kramer}, {Freire}, \&
  {Karuppusamy}}]{Liu2014}
{Liu}, K., {Desvignes}, G., {Cognard}, I., {et~al.} 2014, \mnras, 443, 3752

\bibitem[{Lorimer \& Kramer(2004)}]{LorimerKramer2004_handbook}
Lorimer, D. \& Kramer, M. 2004, {Handbook of pulsar astronomy} (Cambridge
  University Press)

\bibitem[{{Maan} {et~al.}(2020){Maan}, {van Leeuwen}, \& {Vohl}}]{Maan2020}
{Maan}, Y., {van Leeuwen}, J., \& {Vohl}, D. 2020, arXiv e-prints,
  arXiv:2012.11630

\bibitem[{{Madison} {et~al.}(2019){Madison}, {Cordes}, {Arzoumanian},
  {Chatterjee}, {Crowter}, {DeCesar}, {Demorest}, {Dolch}, {Ellis}, {Ferdman},
  {Ferrara}, {Fonseca}, {Gentile}, {Jones}, {Jones}, {Lam}, {Levin}, {Lorimer},
  {Lynch}, {McLaughlin}, {Mingarelli}, {Ng}, {Nice}, {Pennucci}, {Ransom},
  {Ray}, {Spiewak}, {Stairs}, {Stovall}, {Swiggum}, \& {Zhu}}]{madison+19}
{Madison}, D.~R., {Cordes}, J.~M., {Arzoumanian}, Z., {et~al.} 2019, \apj, 872,
  150

\bibitem[{McLaughlin(2013)}]{McLaughlin2013}
McLaughlin, M.~A. 2013, Classical and Quantum Gravity, 30, 224008

\bibitem[{Perera {et~al.}(2019)Perera, DeCesar, Demorest, Kerr, Lentati, Nice,
  Os{\l}owski, Ransom, Keith, Arzoumanian, Bailes, Baker, Bassa, Bhat, Brazier,
  Burgay, Burke-Spolaor, Caballero, Champion, Chatterjee, Chen, Cognard,
  Cordes, Crowter, Dai, Desvignes, Dolch, Ferdman, Ferrara, Fonseca, Goldstein,
  Graikou, Guillemot, Hazboun, Hobbs, Hu, Islo, Janssen, Karuppusamy, Kramer,
  Lam, Lee, Liu, Luo, Lyne, Manchester, McKee, McLaughlin, Mingarelli,
  Parthasarathy, Pennucci, Perrodin, Possenti, Reardon, Russell, Sanidas,
  Sesana, Shaifullah, Shannon, Siemens, Simon, Spiewak, Stairs, Stappers,
  Swiggum, Taylor, Theureau, Tiburzi, Vallisneri, Vecchio, Wang, Zhang, Zhang,
  Zhu, \& Zhu}]{Perera2019}
Perera, B.~B., DeCesar, M.~E., Demorest, P.~B., {et~al.} 2019, Monthly Notices
  of the Royal Astronomical Society, 490, 4666

\bibitem[{{Pesnell} {et~al.}(2012){Pesnell}, {Thompson}, \&
  {Chamberlin}}]{Sdo_paper}
{Pesnell}, W.~D., {Thompson}, B.~J., \& {Chamberlin}, P.~C. 2012, \solphys,
  275, 3

\bibitem[{{Reddy} {et~al.}(2017){Reddy}, {Kudale}, {Gokhale}, {Halagalli},
  {Raskar}, {de}, {Gnanaraj}, {Ajith Kumar}, \& {Gupta}}]{Reddy2017}
{Reddy}, S.~H., {Kudale}, S., {Gokhale}, U., {et~al.} 2017, Journal of
  Astronomical Instrumentation, 6, 1641011

\bibitem[{{Susobhanan} {et~al.}(2021){Susobhanan}, {Maan}, {Joshi}, {Prabu},
  {Desai}, {Nobleson}, {Susarla}, {Girgaonkar}, {Dey}, {Batra}, {Gupta},
  {Gopakumar}, {Bagchi}, {Basu}, {Bethapudi}, {Choudhary}, {De},
  {Krishnakumar}, {Manoharan}, {Naidu}, {Pathak}, {Singha}, \&
  {Surnis}}]{Susobhanan2020}
{Susobhanan}, A., {Maan}, Y., {Joshi}, B.~C., {et~al.} 2021, \pasa, 38, e017

\bibitem[{{Swarup} {et~al.}(1991){Swarup}, {Ananthakrishnan}, {Kapahi}, {Rao},
  {Subrahmanya}, \& {Kulkarni}}]{Swarup1991}
{Swarup}, G., {Ananthakrishnan}, S., {Kapahi}, V.~K., {et~al.} 1991, Current
  Science, 60, 95

\bibitem[{{Taylor}(1992)}]{Taylor92}
{Taylor}, J.~H. 1992, Philosophical Transactions of the Royal Society of London
  Series A, 341, 117

\bibitem[{{Tiburzi} {et~al.}(2021){Tiburzi}, {Shaifullah}, {Bassa}, {Zucca},
  {Verbiest}, {Porayko}, {van der Wateren}, {Fallows}, {Main}, {Janssen},
  {Anderson}, {Bak Nielsen}, {Donner}, {Keane}, {K{\"u}nsem{\"o}ller},
  {Os{\l}owski}, {Grie{\ss}meier}, {Serylak}, {Br{\"u}ggen}, {Ciardi},
  {Dettmar}, {Hoeft}, {Kramer}, {Mann}, \& {Vocks}}]{Tiburzi2020}
{Tiburzi}, C., {Shaifullah}, G.~M., {Bassa}, C.~G., {et~al.} 2021, \aap, 647,
  A84

\bibitem[{{Tiburzi} {et~al.}(2019){Tiburzi}, {Verbiest}, {Shaifullah},
  {Janssen}, {Anderson}, {Horneffer}, {K{\"u}nsem{\"o}ller}, {Os{\l}owski},
  {Donner}, {Kramer}, {Kumari}, {Porayko}, {Zucca}, {Ciardi}, {Dettmar},
  {Grie{\ss}meier}, {Hoeft}, \& {Serylak}}]{Tiburzi2019}
{Tiburzi}, C., {Verbiest}, J.~P.~W., {Shaifullah}, G.~M., {et~al.} 2019,
  \mnras, 487, 394

\bibitem[{{van Straten} \& {Bailes}(2011)}]{vanStraten2011}
{van Straten}, W. \& {Bailes}, M. 2011, \pasa, 28, 1

\bibitem[{{Verbiest} {et~al.}(2016){Verbiest}, {Lentati}, {Hobbs}, {van
  Haasteren}, {Demorest}, {Janssen}, {Wang}, {Desvignes}, {Caballero}, {Keith},
  {Champion}, {Arzoumanian}, {Babak}, {Bassa}, {Bhat}, {Brazier}, {Brem},
  {Burgay}, {Burke-Spolaor}, {Chamberlin}, {Chatterjee}, {Christy}, {Cognard},
  {Cordes}, {Dai}, {Dolch}, {Ellis}, {Ferdman}, {Fonseca}, {Gair},
  {Garver-Daniels}, {Gentile}, {Gonzalez}, {Graikou}, {Guillemot}, {Hessels},
  {Jones}, {Karuppusamy}, {Kerr}, {Kramer}, {Lam}, {Lasky}, {Lassus},
  {Lazarus}, {Lazio}, {Lee}, {Levin}, {Liu}, {Lynch}, {Lyne}, {Mckee},
  {McLaughlin}, {McWilliams}, {Madison}, {Manchester}, {Mingarelli}, {Nice},
  {Os{\l}owski}, {Palliyaguru}, {Pennucci}, {Perera}, {Perrodin}, {Possenti},
  {Petiteau}, {Ransom}, {Reardon}, {Rosado}, {Sanidas}, {Sesana}, {Shaifullah},
  {Shannon}, {Siemens}, {Simon}, {Smits}, {Spiewak}, {Stairs}, {Stappers},
  {Stinebring}, {Stovall}, {Swiggum}, {Taylor}, {Theureau}, {Tiburzi},
  {Toomey}, {Vallisneri}, {van Straten}, {Vecchio}, {Wang}, {Wen}, {You},
  {Zhu}, \& {Zhu}}]{Verbiest2016}
{Verbiest}, J.~P.~W., {Lentati}, L., {Hobbs}, G., {et~al.} 2016, \mnras, 458,
  1267

\bibitem[{{You} {et~al.}(2007){You}, {Hobbs}, {Coles}, {Manchester}, \&
  {Han}}]{You2007}
{You}, X.~P., {Hobbs}, G.~B., {Coles}, W.~A., {Manchester}, R.~N., \& {Han},
  J.~L. 2007, \apj, 671, 907

\end{thebibliography}

\end{document}